\documentclass[aps,twocolumn,showpacs,amsmath,amssymb,superscriptaddress]{revtex4}
\usepackage{epsf}
\usepackage{bm}
\usepackage{multirow}
\usepackage[dvips]{graphicx}
\usepackage{dcolumn}

\newcommand{\etal}{{ \em et al }}

\begin{document}
\title[]{A Systematic Study of Electronic Structure from
Graphene to {\it Graphane}}

\author{Prachi Chandrachud}
\affiliation{Department of Physics, University of Pune, Ganeshkhind, Pune--411 007,
India.}
 
\author{Bhalchandra S Pujari}
\altaffiliation[Present address: ]{National Institute of Nanotechnology, 11421
Saskatchewan Drive, Edmonton, Alberta, T6G 2M9, Canada}

\affiliation{Department of Physics, University of Pune, Ganeshkhind, Pune--411 007,
India.} 

\author{Soumyajyoti Haldar}
\email[Corresponding Author: ]{shaldar@cms.unipune.ac.in}
\affiliation{Department of Physics, University of Pune, Ganeshkhind, Pune--411 007,
India.} 
\affiliation{Centre for Modelling and Simulations, University of Pune,  Ganeshkhind,
Pune--411 007, India.}

\author{Biplab Sanyal} 
\affiliation{Department of Physics and Astronomy, Uppsala University, Box 516, 75120
Uppsala, Sweden} 

\author{D G Kanhere}
\affiliation{Department of Physics, University of Pune, Ganeshkhind, Pune--411 007,
India.} 
\affiliation{Centre for Modelling and Simulations, University of Pune,  Ganeshkhind,
Pune--411 007, India.}

\begin{abstract}

While graphene is a semi-metal, a recently synthesized hydrogenated graphene called {\it
graphane}, is an insulator.  We have probed the transformation of graphene upon
hydrogenation to {\it graphane} within the framework of density functional theory. By
analyzing the electronic structure for eighteen different hydrogen concentrations, we
bring out some novel features of this transition. Our results show that the hydrogenation
favours clustered configurations leading to the formation of compact islands. The analysis
of the charge density and electron localization function (ELF) indicates that as hydrogen
coverage increases the semi-metal turns into a metal showing a delocalized charge density,
then transforms into an insulator. The metallic phase is spatially inhomogeneous in the
sense, it contains the islands of insulating regions formed by hydrogenated carbon atoms
and the metallic channels formed by contiguous bare carbon atoms. It turns out that it is
possible to pattern the graphene sheet to tune the electronic structure. For example
removal of hydrogen atoms along the diagonal of the unit cell yielding  an armchair
pattern  at the edge gives rise to a band gap of 1.4 eV. We also show that a weak
ferromagnetic state exists even for a large hydrogen coverage whenever there is a
sublattice imbalance in presence of odd number of hydrogen atoms.

\end{abstract}

\maketitle
\section{Introduction}

Carbon is regarded as one of the most versatile elements in the periodic table forming a
wide variety of structures such as three dimensional sp$^{3}$ bonded solids like diamond,
sp$^{2}$ hybridized two dimensional systems like graphene and novel nano structures like
fullerenes and nanotubes.  The electronic structure and the physical properties of these
carbon based materials are turning out to be exotic \cite{smalley}. Although the existence
and the properties of three dimensional allotrope, graphite containing weakly coupled
stacks of graphene layers were well known \cite{wallace}, the experimental realization of
a monolayer graphene, brought forth a completely different set of novel properties
\cite{gaim, castro-phys}. The triangular bipartite lattice of graphene leads to the
electronic structure having a linear dispersion near the Dirac points.  The low energy
behaviour of such two dimensional electrons in graphene has been a subject of intense
experimental and theoretical activity exploring the electronic, magnetic, mechanical,
transport properties etc.  For a recent review the reader is referred to Castro Neto {\it
et. al.} ~\cite{castro}.

Although graphene is considered as a prime candidate for many applications, the absence of
a band gap is a worrisome feature for the applications to the solid state electronic
devices. In the past, several routes have been proposed to open a band gap
\cite{route1,route2,route3}.  The most interesting one is the recent discovery of a
completely hydrogenated graphene sheet named as {\it graphane}.  {\it Graphane} was first
predicted by Sofo {\it et.  al.}~\cite{sofo} on the basis of electronic structure
calculations and has been recently synthesized by Elias {\it et. al.} ~\cite{elias}. The
experimental work also showed that the process of hydrogenation is reversible,  making
{\it graphane} a potential candidate for hydrogen storage systems.  Since upon
hydrogenation, graphene, a semi-metal turns into an insulator, it is a good candidate for
investigating the nature of metal insulator transition (MIT). 

There are few reports investigating the electronic structure of graphene sheets as a
function of hydrogen coverage leading to the opening of a bandgap. Most of these are
restricted to a small number of hydrogen atoms. The electronic structure of hydrogen
adsorbed on graphene has been investigated using density functional theory (DFT) by
Boukhvalov {\it et. al.}\cite{boukhvalov} and Casolo {\it et. al.} \cite{casolo}. Their
results support the use of graphene as a hydrogen storage material. Their work also shows
that the thermodynamically and kinetically favoured structures are those that minimize the
sublattice imbalance.  Flores {\it et. al.} have investigated the role of
hydrogen-frustration in {\it graphane} like structures using ab-initio methods and
reactive classical molecular dynamics \cite{flores}. The stability trends in small
clusters of hydrogen on graphene have been discussed in a recent review \cite{roman}. A
recent work by Zhou {\it et. al.} \cite{zhou} predicted  a new ordered ferromagnetic state
obtained by removing hydrogen atoms from one side of the plane of {\it graphane}.
Recently, Wu {\it et. al.} \cite{Wu} have reported implications of selective hydrogenation
by designing an array of triangular carbon domains separated by hydrogenated strips. The
electronic structure of the interface between graphene and the hydrogenated part has been
investigated by Schmidt and Loss \cite{sc-loss}. They show the existence of edge states
for zigzag interface having strong spin orbit interaction. An interesting experimental
work using scanning tunnelling microscopy, by Hornek\ae{}r {\it et.al}
\cite{horneka1,horneka2} demonstrates clustering of hydrogen atoms on graphite surface. The work
shows that there is vanishingly small adsorption barrier for hydrogen in the vicinity of
already adsorbed hydrogen atoms. The question of the effect of defect on the electronic structure
of atomic hydrogen has been addressed by Duplock {\it et.al} \cite{dup} within the density
functional theory. They show that the electronic gap state associated with the adsorbed
hydrogen is very sensitive to the presence of defect such as Stone-Wales defect.  The
localization behaviour of disordered graphene by hydrogenation has been reported within
the tight binding formalism \cite{bang}. The potential of hydrogenated graphene
nanoribbons for spintronics applications has been investigated within DFT, by Soriano {\it
et.al} \cite{soriano}. The defect and disorder induced magnetism in graphene (including
adsorbed hydrogen as a defect) has been studied by Yazyev, Yazyev and Helm mostly using
Hubbard Hamiltonian and DFT \cite{ya1,ya2,ya3}. Theoretical investigations have also been
carried out for a single hydrogen defect on {\it graphane} sheet using GW method
\cite{GW}, and for one and two vacancies in {\it graphane} \cite{pujari} using DFT. It has
been shown that  interaction between adatoms in hydrogenated graphene is long range and
its nature is dependant on which sublattice the adatoms reside \cite{shy}. The result
suggests that the adatoms tend to aggregate. A very recent work explores a formation of
quantum dots as small island of graphene in {\it graphane} host \cite{singh-nm}.

In the present work, we investigate some aspects of graphene - {\it graphane} transition
by probing the electronic structure of hydrogenated graphene. The objective of our work is
to understand the evolution of the electronic structure upon hydrogenation and gain some
insight into the way band gap opens.  Therefore we have carried out extensive calculations
for eighteen different hydrogen coverages between graphene (0\% hydrogen coverage) and
{\it graphane} (100\% hydrogen coverage) within the framework of DFT.  Our results suggest
that the hydrogenation in graphene takes place via clustering of hydrogens. Analysis based
on density of states (DOS) indicates that before the gap opens (as a function of hydrogen
coverage) the hydrogenated graphene sheet acquires a metallic character. This metallic
state is spatially inhomogeneous in the sense, it consists of insulating regions of
hydrogenated carbon atoms surrounded by the conducting channels formed by the bare carbon
atoms. We also show that it is possible to tune the electronic structure by the selective
decoration of hydrogen atoms to achieve semiconducting or metallic state.

In the next section we present the computational details, followed by results and
discussions.

\section{Computational Details}

The calculations have been performed using a plane-wave projector augmented wave method
based code, VASP~\cite{vasp}.  The generalized gradient approximation as proposed by
Perdew, Burke and Ernzerhof \cite{gga2} has been used for the exchange-correlation
potential. The convergence of binding energies with respect to the size of the supercell
has been checked using three different sizes, viz., 5 $\times$ 5, 6 $\times$ 6 and 7
$\times$ 7 containing 50, 72 and 98 carbon atoms for the case of 50\% hydrogen coverage.
The binding energy per atom changes by about 0.05 eV (a percentage change of 0.06\%) in
going from 5 $\times$ 5 to 7 $\times$ 7.  Therefore we have chosen a 5 $\times$ 5 unit
cell for the coverage upto 50\% of hydrogen and 6 $\times$ 6 for the higher coverages.
This choice is consistent with the one used by Leb{\`e}gue {\it et.al} \cite{GW}. In order
to obtain adequate convergence in the density of states, we have carried out calculations
on different {\bf $k$} grids. It was found that at least 9 $\times$ 9 {\bf $k$} grid was
required during geometry optimization for an acceptable convergence. However, a minimum of
17 $\times$ 17 {\bf $k$} grid was necessary for obtaining  an accurate DOS.  The
convergence criterion used for the total energy and the force are 10$^{-5}$ eV and 0.005
eV/{\AA} respectively.  

All the calculations have been performed on the chair conformer configuration where
hydrogen atoms are attached to carbon atoms alternatively on opposite sides of the plane.
This is known to be a lower energy configuration as compared to the boat conformer
\cite{sofo,GW}.

\section{Results and discussion}

\begin{table}
\begin{center}

\caption{\label{tab:1}Shifted binding energies (BE) for different configuration of 20\%
coverage (10H + 50C). The binding energy of the most stable structure is the
reference(zero) level.  The more negative binding energy means more unstable structure.\\}

\begin{tabular}{|p{5cm}|c|} 
\hline 
\multicolumn{2}{|c|}{20\% Hydrogen Coverage} \\ 
\hline 
Configuration & BE (total cell) (eV)\\ \hline 
Random placement of hydrogens & -9.32\\ \hline
Five hydrogen pairs placed separately & -4.29\\ \hline
Hydrogens placed along the diagonal of the cell & -2.85\\ \hline
Two clusters of seven and three hydrogens & -2.15\\ \hline
A single hydrogen isolated from the cluster of nine hydrogens & -1.77\\ \hline
One compact cluster & 0.0\\ \hline
\end{tabular}
\end{center}
\end{table}

In order to decide the minimum energy positions for hydrogen atoms, following procedure
has been adopted. Upto 20\% coverage of hydrogen, we have carried out the geometry
optimization for two different configurations of hydrogen atoms, first by placing the
hydrogen atoms randomly and second by placing the hydrogen atoms contiguously, so as to
form a compact cluster of hydrogenated carbon atoms. It turns out that, in all the cases,
the configuration forming the compact cluster  of the hydrogen atoms is energetically
favoured. In order to assess the relative stability of different patterns  we have
calculated six different patterns for 20\% coverage case as shown in table \ref{tab:1}.
The binding energy of the most stable structure is the reference level. The more negative
binding energy means more unstable structure.  

\begin{figure*}
\begin{center}
\includegraphics[width=5.5cm]{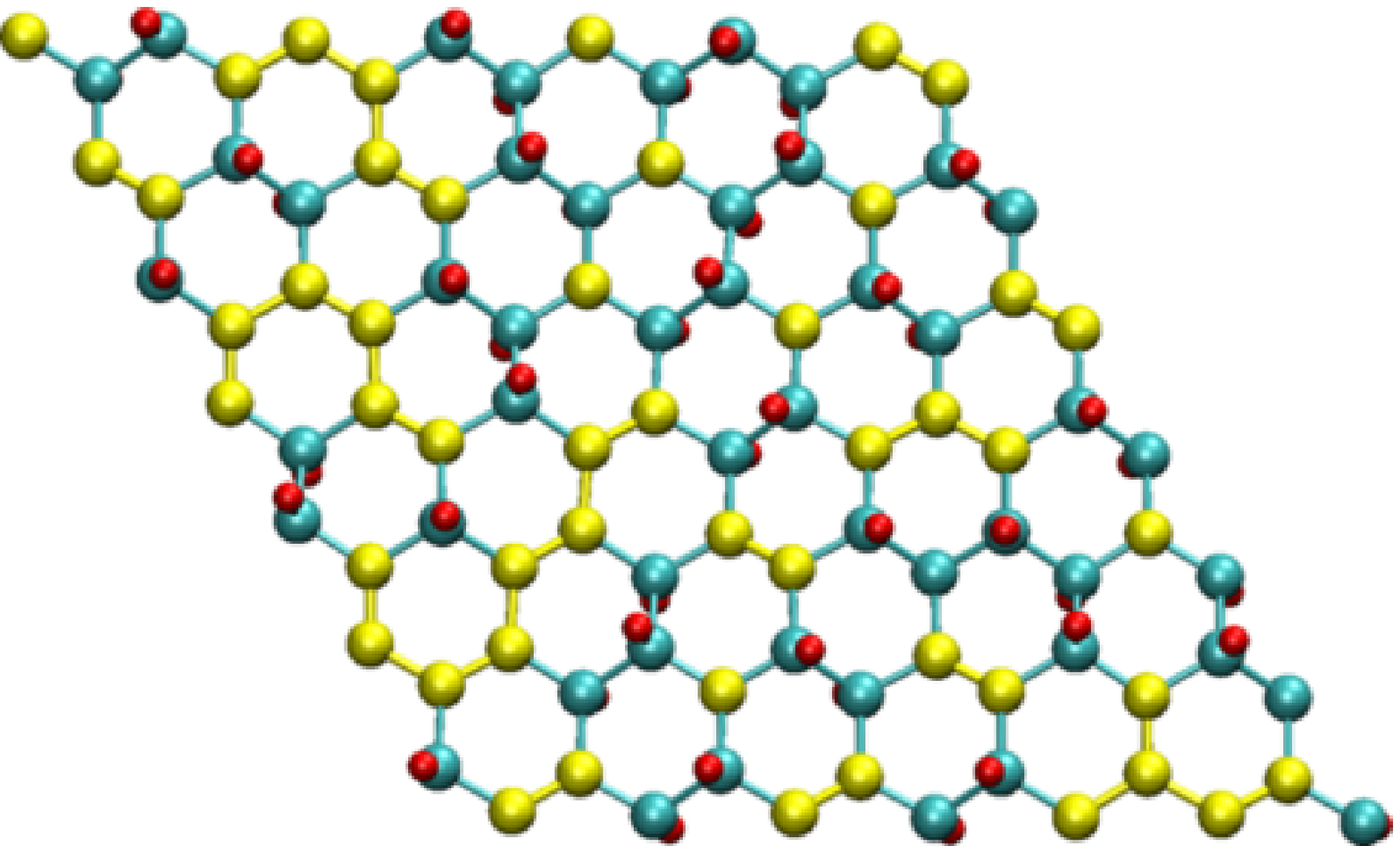}
\includegraphics[width=5.5cm]{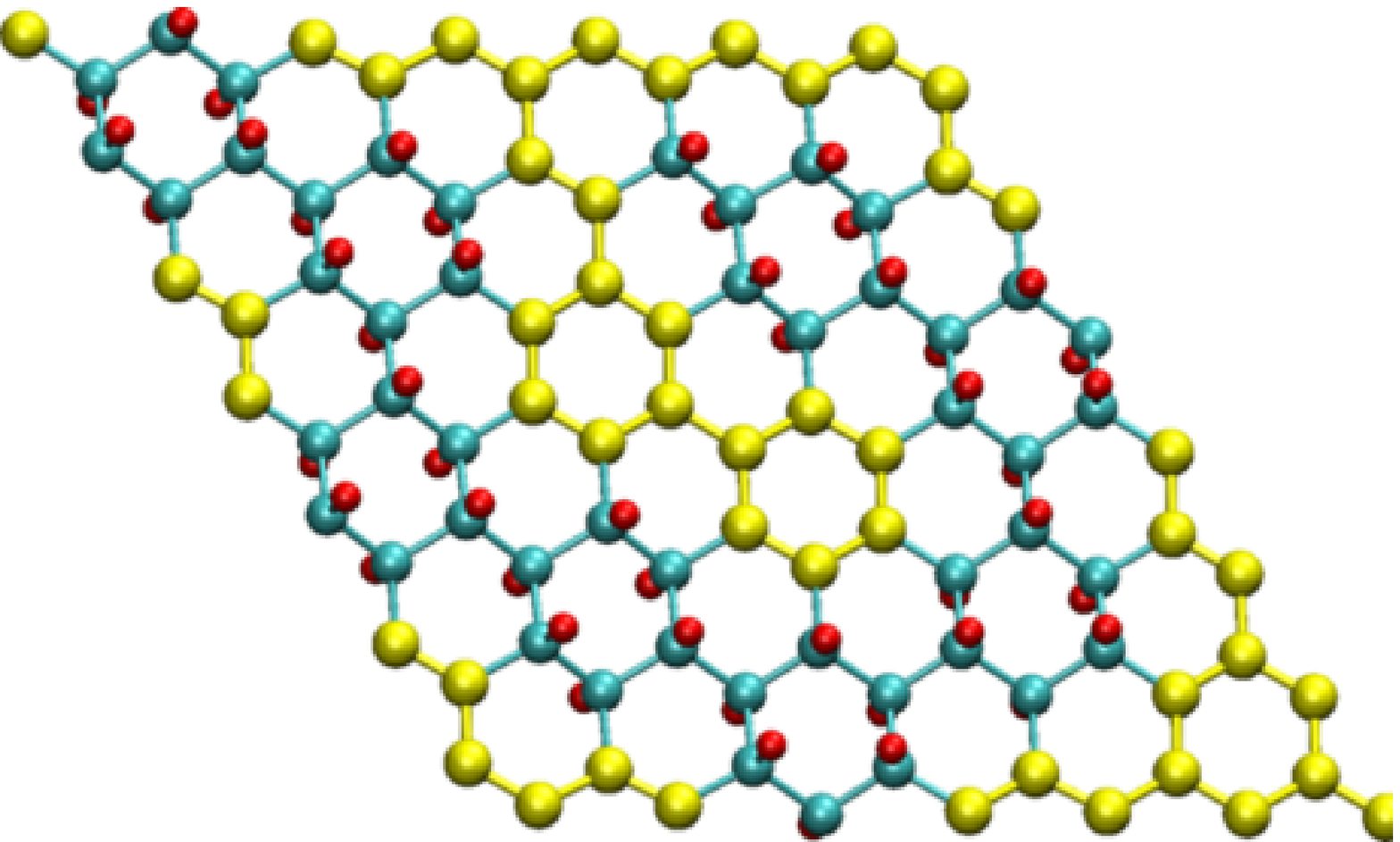}\\
(a) \hskip 5cm (b) \\
\includegraphics[width=5.5cm]{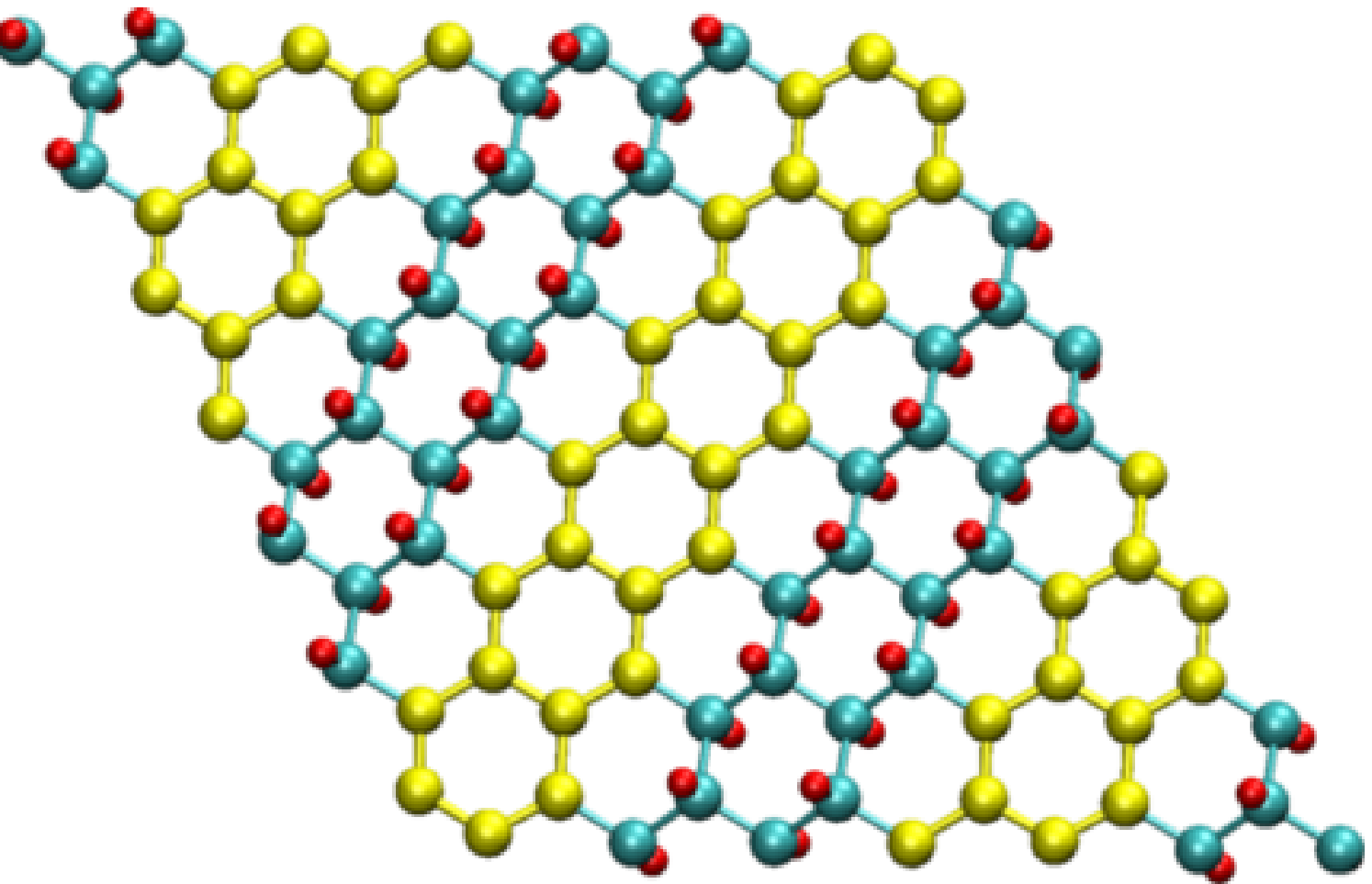} 
\includegraphics[width=5.5cm]{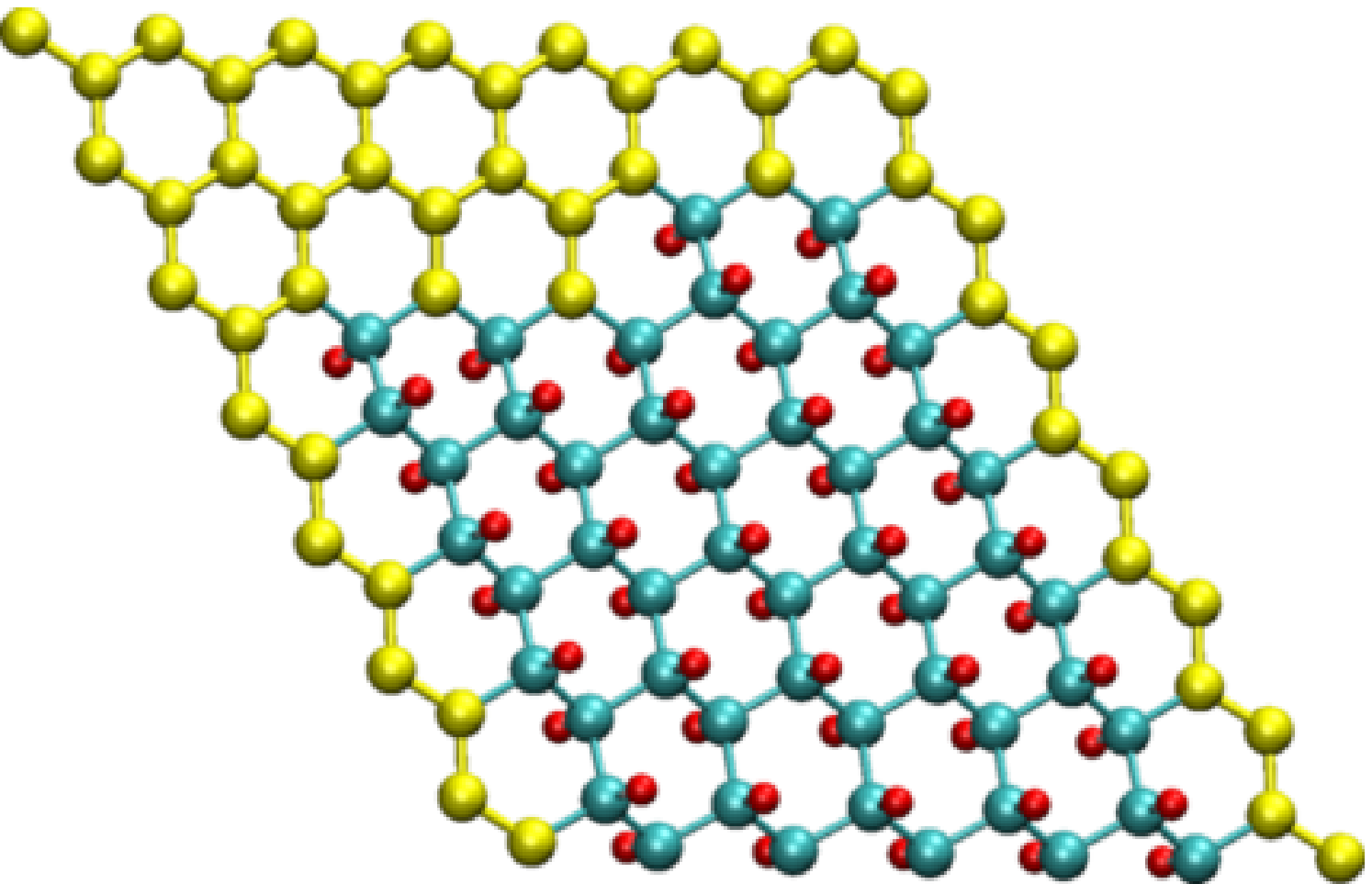} \\
(c) \hskip 5cm (d) \\
\includegraphics[width=5.5cm]{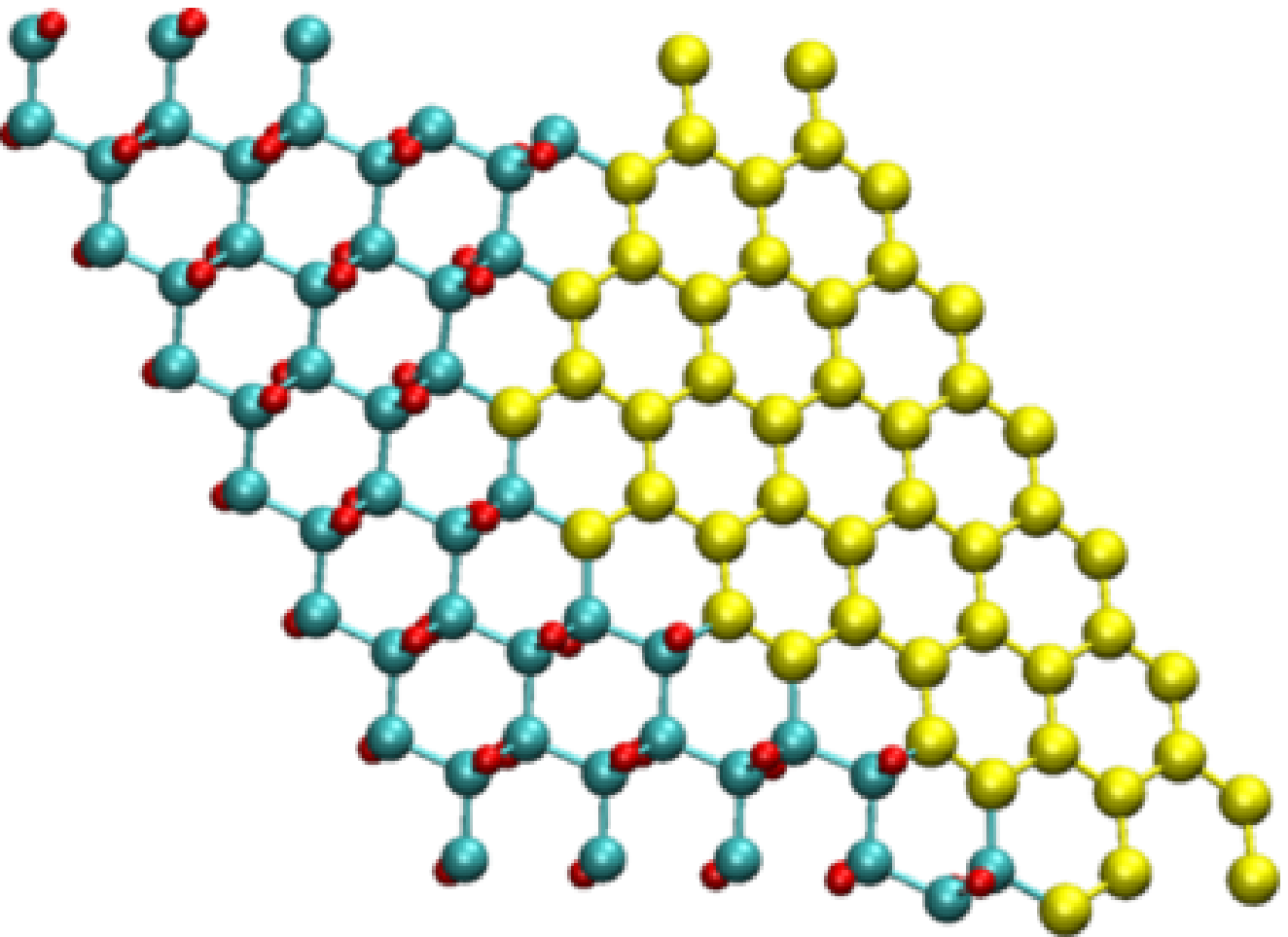} 
\includegraphics[width=5.5cm]{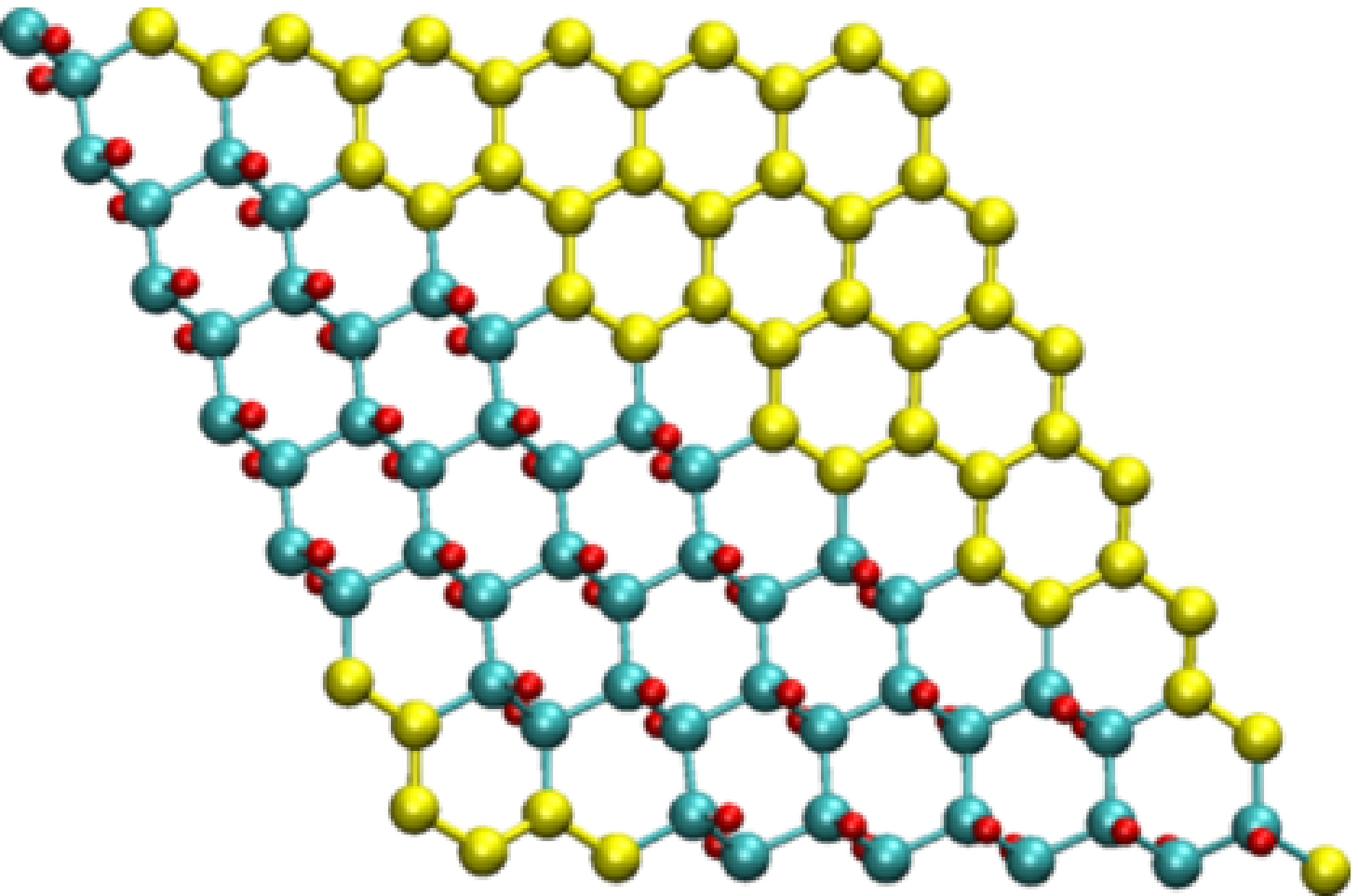} \\
(e) \hskip 5cm (f) \\
\includegraphics[width=5.5cm]{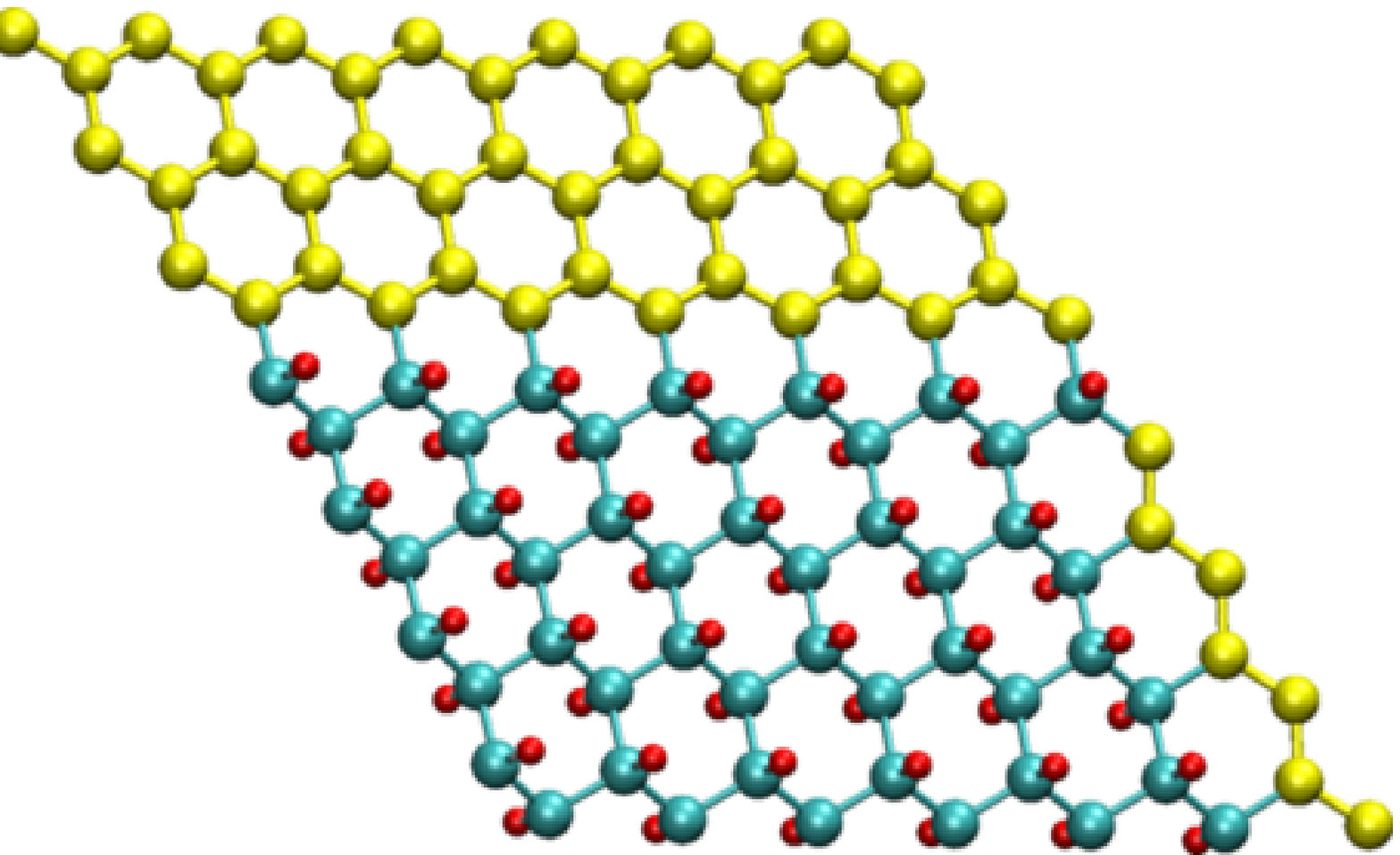}\\
(g)
\end{center}

\caption{\label{fig1} (Colour online) Different hydrogen decorations for 50\%
concentration.  In figure, yellow (in print light shaded) balls are bare carbon atoms,
turquoise (in print darker shades) balls are hydrogenated carbons and red (in print dark
small) balls are hydrogen atoms.}

\end{figure*}

We note that the compact cluster configuration is the lowest in energy and is lower by 0.9
eV/H atom compared to the energy of the configuration with randomly placed hydrogen atoms.
Thus our geometry optimization shows a preference for hydrogens to decorate the graphene
lattice in a contiguous and compact manner. In order to assess  the validity of this
process for the larger coverages and to understand the influence of different edge
patterns like zigzag and arm chair, we have considered seven different patterns for 50\%
coverage,  as shown in figure \ref{fig1}. These calculations have been carried out on a
larger unit cell containing 98 atoms of carbon.  Although a compact cluster configuration
is energetically preferred one, there are different ways in which  the hydrogens can be
placed yielding a compact geometry. The seven cases considered are : 1) Randomly
distributed hydrogens (figure \ref{fig1}(a)), 2) A chain of bare carbon atoms (figure
\ref{fig1}(b)), 3) Three separated clusters as shown in (figure \ref{fig1}(c)), 4) A
single cluster having zigzag interface and with one line of bare carbon atoms at the two
side edges (figure \ref{fig1}(d)), 5) A mixture of armchair and zigzag pattern at the
interface (figure \ref{fig1}(e)), 6) A armchair pattern at the interface (figure
\ref{fig1}(f)) and 7) A zigzag pattern at the interface (figure \ref{fig1}(g)). 

\begin{table}
\begin{center}

\caption{\label{tab:2}Shifted binding energies (BE) for different configuration of 50\%
coverage as shown in figure \ref{fig1}. The binding energy of the most stable structure is
the reference(zero) level. The more negative binding energy means more unstable
structure.\\}

\begin{tabular}{|p{5cm}|c|} 
\hline 
\multicolumn{2}{|c|}{50\% Hydrogen Coverage on 98 Carbon cell} \\ 
\hline 
Configuration & BE (total cell) (eV)\\ \hline
Random placement of hydrogens (figure \ref{fig1}(a)) & -23.29\\ \hline
Chain of connecting bare Carbon atoms (figure \ref{fig1}(b)) & -3.65 \\ \hline
Three separated islands (figure \ref{fig1}(c)) & -3.25\\ \hline

Zigzag interface with a line of bare carbon atoms at the two side edges (figure
\ref{fig1}(d)) & -2.46\\ \hline 

Compact mixed (zigzag \& armchair) interface (figure \ref{fig1}(e)) & -1.66 \\ \hline
Compact armchair interface (figure \ref{fig1}(f)) & -1.43 \\ \hline
Compact zigzag interface (figure \ref{fig1}(g)) & 0.0 \\ \hline 
\end{tabular}
\end{center}
\end{table}

In the zigzag case each hexagonal ring at the edge has three hydrogenated carbons and
three empty carbons. An armchair pattern consists of only two hydrogenated carbons at the
edge. The binding energies for all these seven structures (shown in figure \ref{fig1}) are
compared in the table \ref{tab:2}. Even among the islands, the `most compact' one (having
minimum surface area of covered hydrogens) has the highest binding energy. By the most
compact, we mean an island of hydrogenated carbons which covers the minimum area. We have
discussed more details about the zigzag and arm chair patterned configurations in section
\ref{subsec:tune}. 

The tendency to form the compact cluster can be understood by noting that the hydrogen
atom placed on the top of an bare carbon atom pulls up the carbon atom above the plane
by 0.33 \AA~and deforms the surrounding lattice points also. Therefore it costs less
energy to place an extra hydrogen atom nearest to the existing cluster, since the number
of strained bonds is  less compared to the case where the hydrogen atom is placed away
from the cluster.  In the latter case the entire neighbourhood is deformed. Our results
are consistent with the work of Hornek\ae{}r \etal~\cite{horneka1}.

\begin{table}
\begin{center}

\caption{\label{tab:3}Shifted binding energies (BE) (eV) and the magnetic
moments($\mu_{B}$) of 4 hydrogen vacancies. The binding energy of the most stable
structure is the reference(zero) level.  The more negative binding energy means more
unstable structure.\\}

\begin{tabular}{|p{3cm}|c|c|}
\hline
\multicolumn{3}{|c|}{92\% Hydrogen Coverage (4 vacancies)}\\ \hline
Configuration & BE (total cell) & Magnetic Moment\\ \hline
Random & -5.50 & 4.0\\ \hline
3H on one hexagon and 1 on other & -2.09 & 1.82\\ \hline
2H pairs placed on different hexagons & -0.11 & nil\\ \hline
Compact single island & 0.0 & nil\\ \hline
\end{tabular}
\end{center}
\end{table}

We have also carried out a similar analysis for the case of four hydrogen vacancies in
{\it graphane}. The calculations for various configurations have been carried out with
spin polarization. The calculated binding energies and the magnetic moments are tabulated
in table \ref{tab:3}. The four configurations considered are 1) 4 hydrogen atoms removed
randomly, 2) 3 hydrogen atoms removed from one hexagon and 1 from other, 3) 2 hydrogen
atom pairs removed from different hexagons and 4) 4 hydrogens removed from a single
hexagon (compact). We observe that the structure having a compact form of vacancies has
the highest binding energy. The magnetic moment is about 2 for the lattice imbalance case.
It is fruitful to recall Lieb's theorem which states that for a bipartite lattice (one
electron per site) the spin S of the ground state is  1/2 $\times$ (the lattice
imbalance)~\cite{lieb}. The magnetic moment seen in the case of random placement is just
the sum of the isolated moments of hydrogen atoms, essentially a non interacting case.

\begin{figure*}
  \begin{center}
    \includegraphics[width=5.6cm,height=3.5cm]{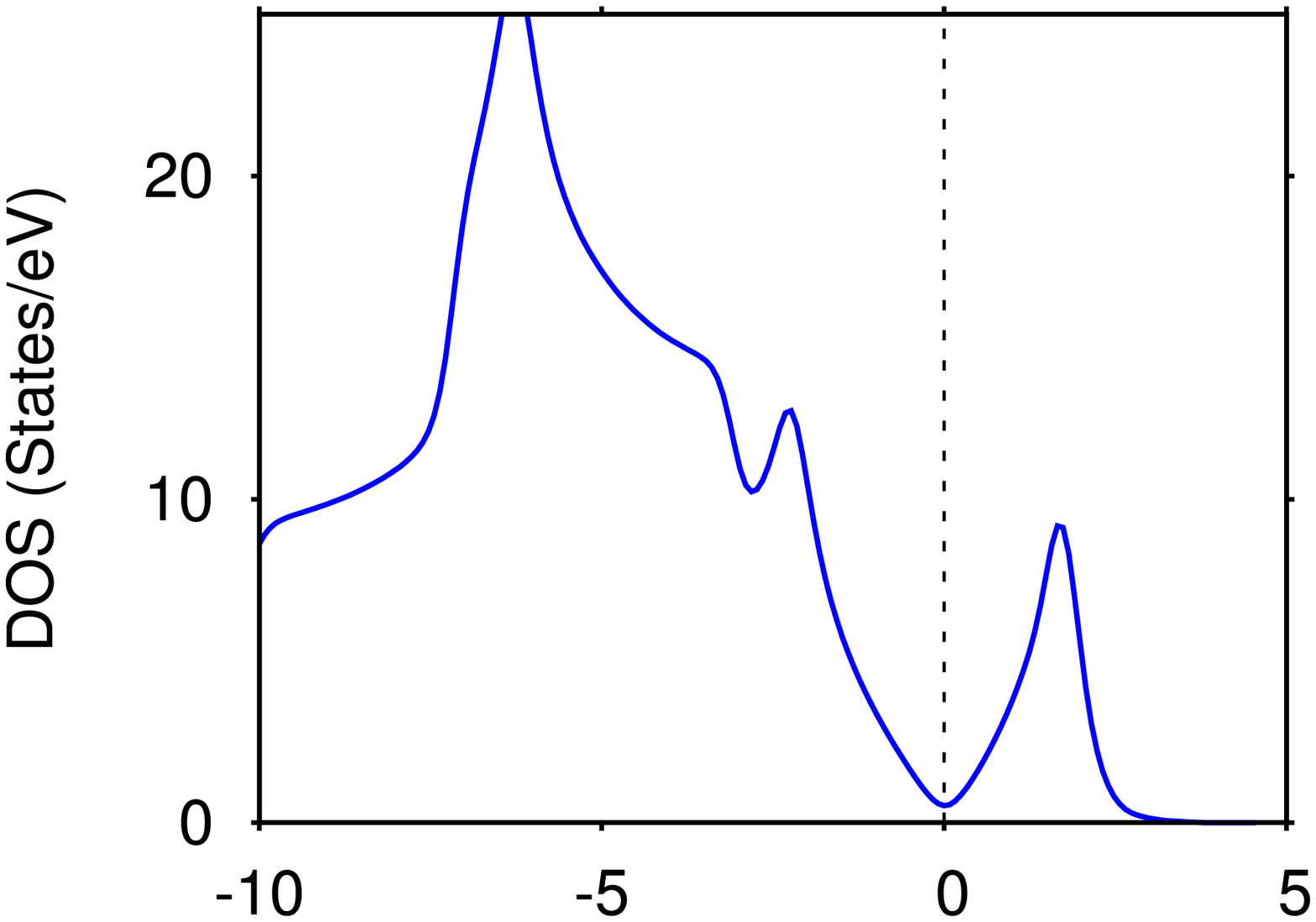}
    \includegraphics[width=5.6cm,height=3.5cm]{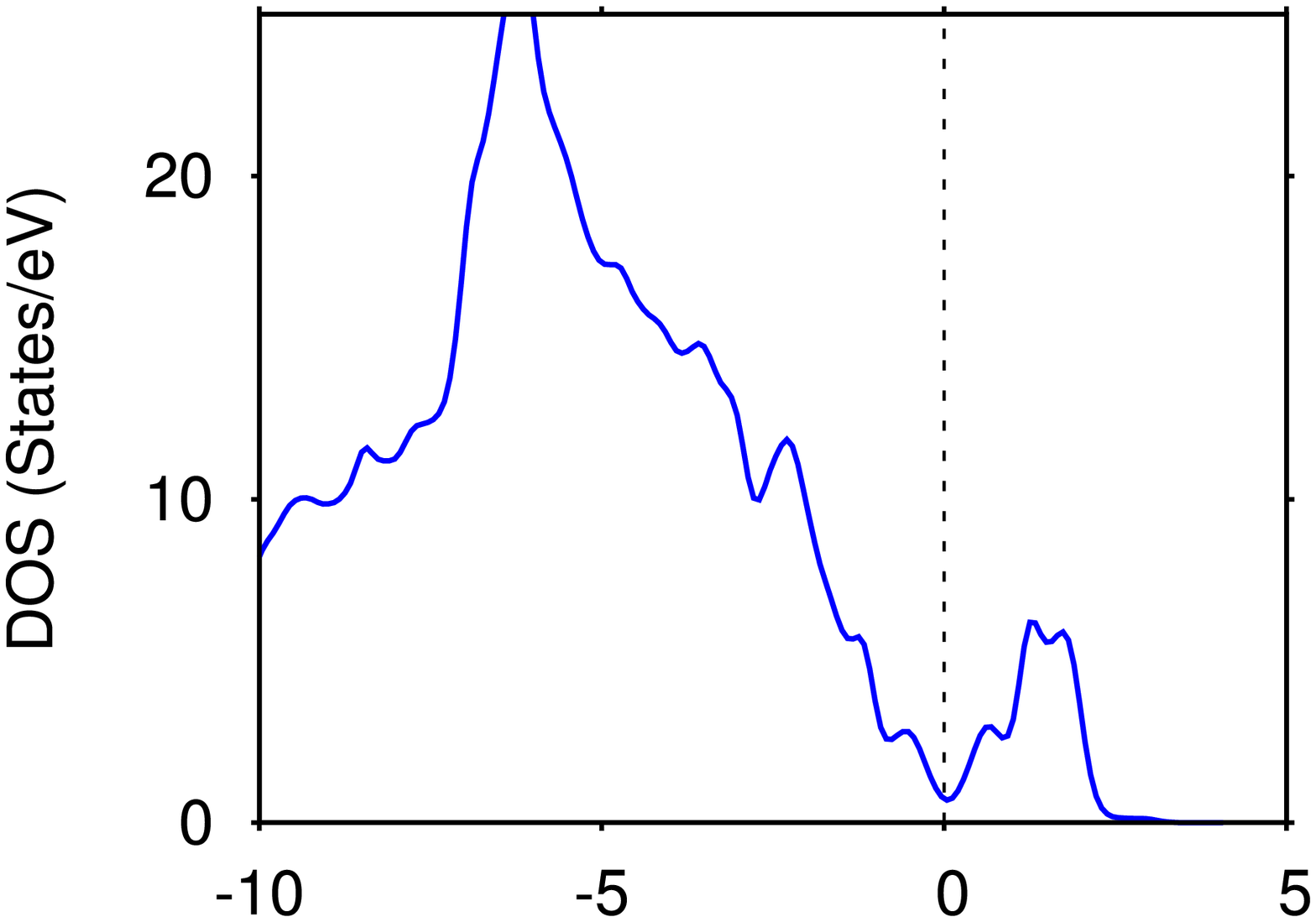}\\
   (a) Graphene \hskip 3cm (b) 4\% hydrogen\\~\\
    \includegraphics[width=5.6cm,height=3.5cm]{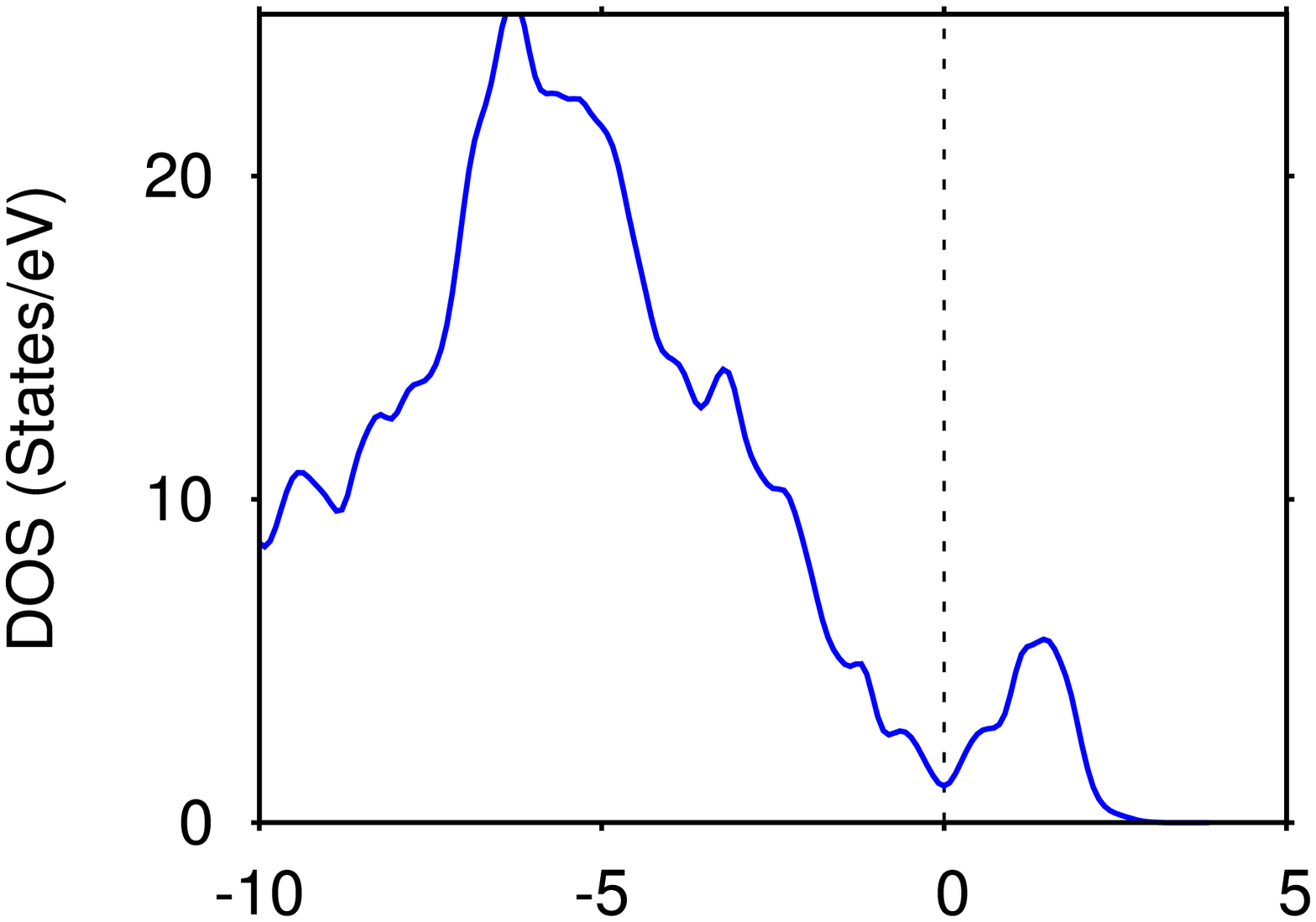}
    \includegraphics[width=5.6cm,height=3.5cm]{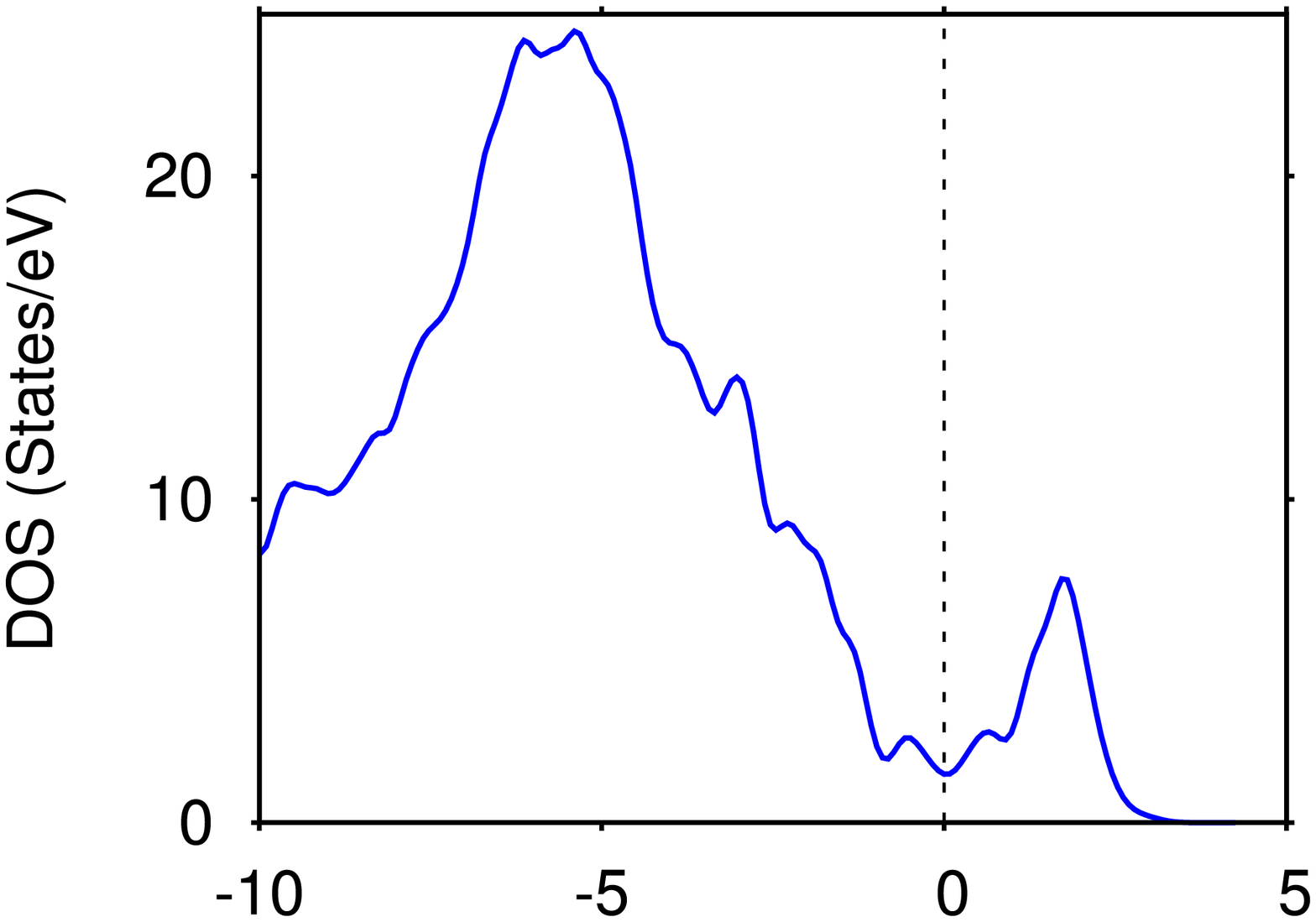}\\
    (c) 16\% hydrogen \hskip 3cm (d) 20\% hydrogen\\~\\
    \includegraphics[width=5.5cm,height=3.3cm]{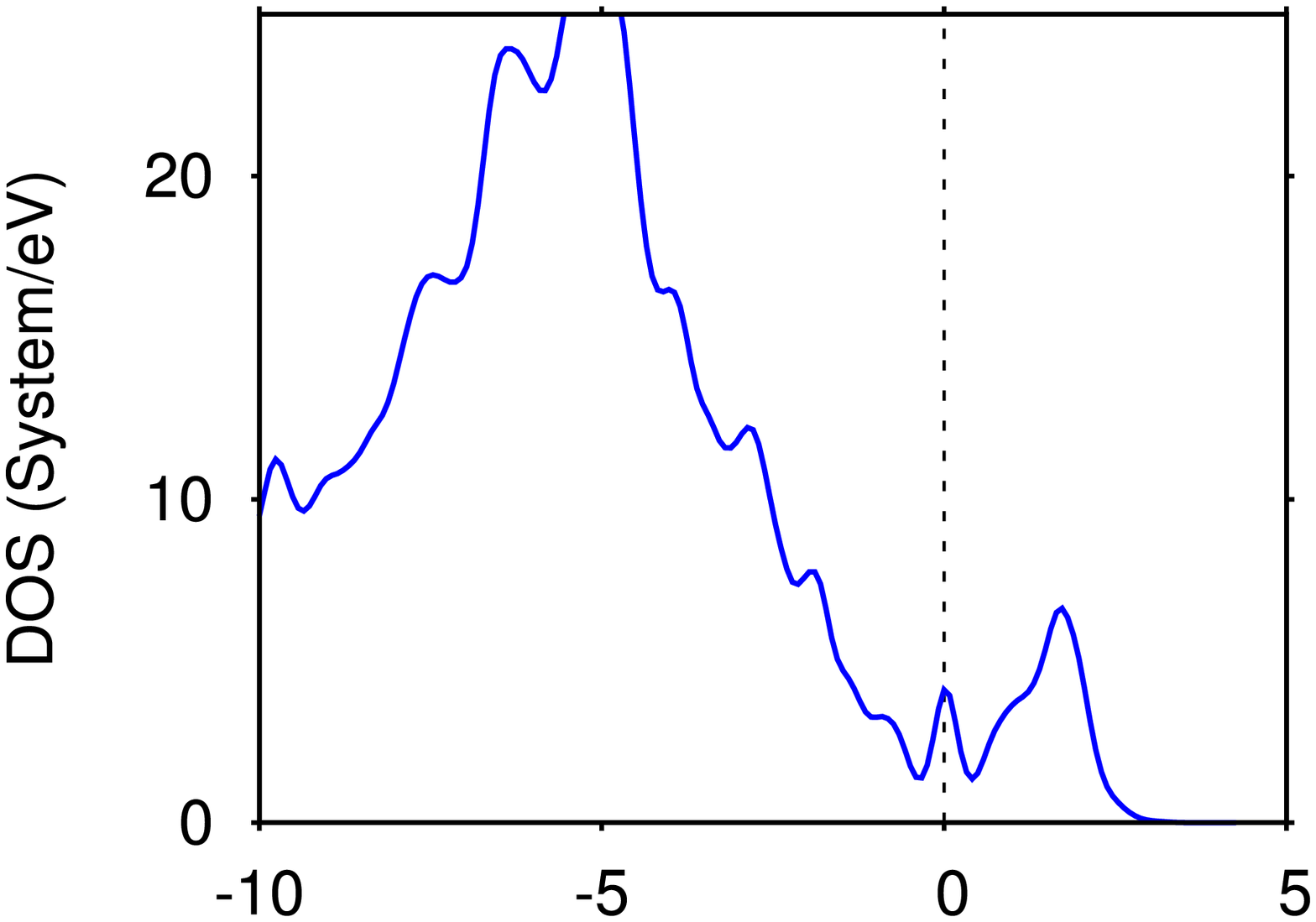}
    \includegraphics[width=5.8cm,height=3.5cm]{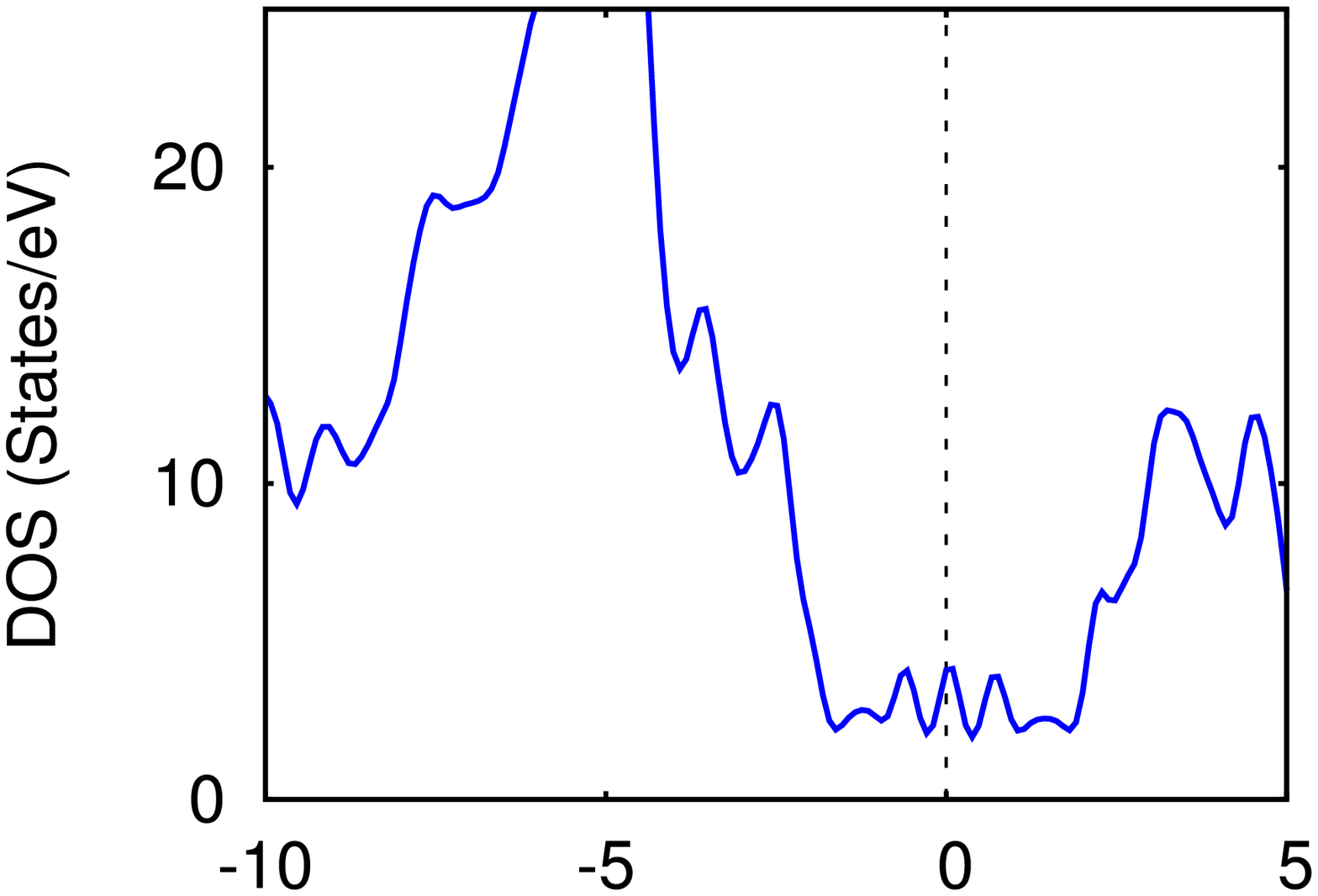}\\
    (e) 30\% hydrogen \hskip 3cm (f) 50\% hydrogen\\~\\
    \includegraphics[width=5.8cm,height=3.5cm]{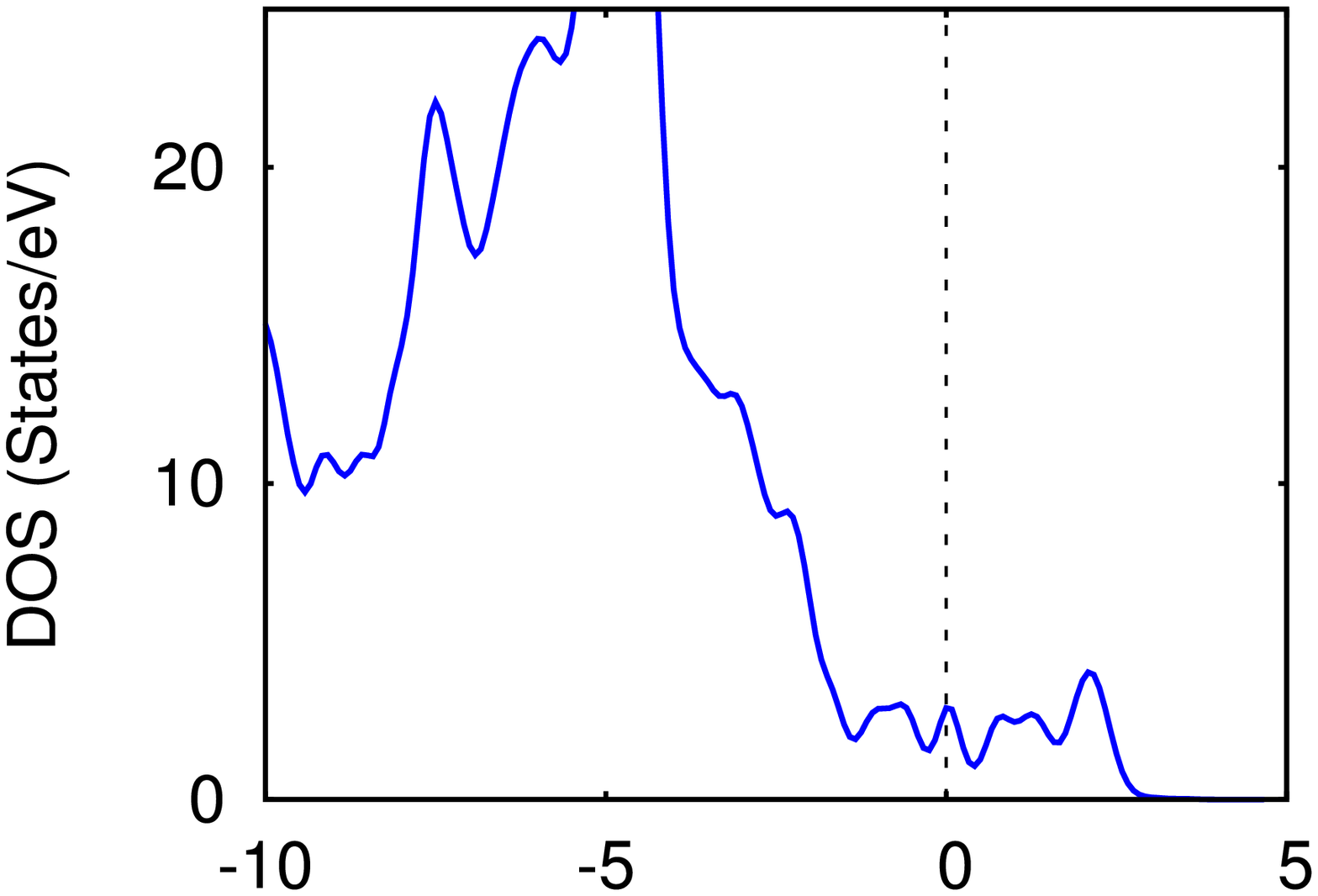}
    \includegraphics[width=5.8cm,height=3.5cm]{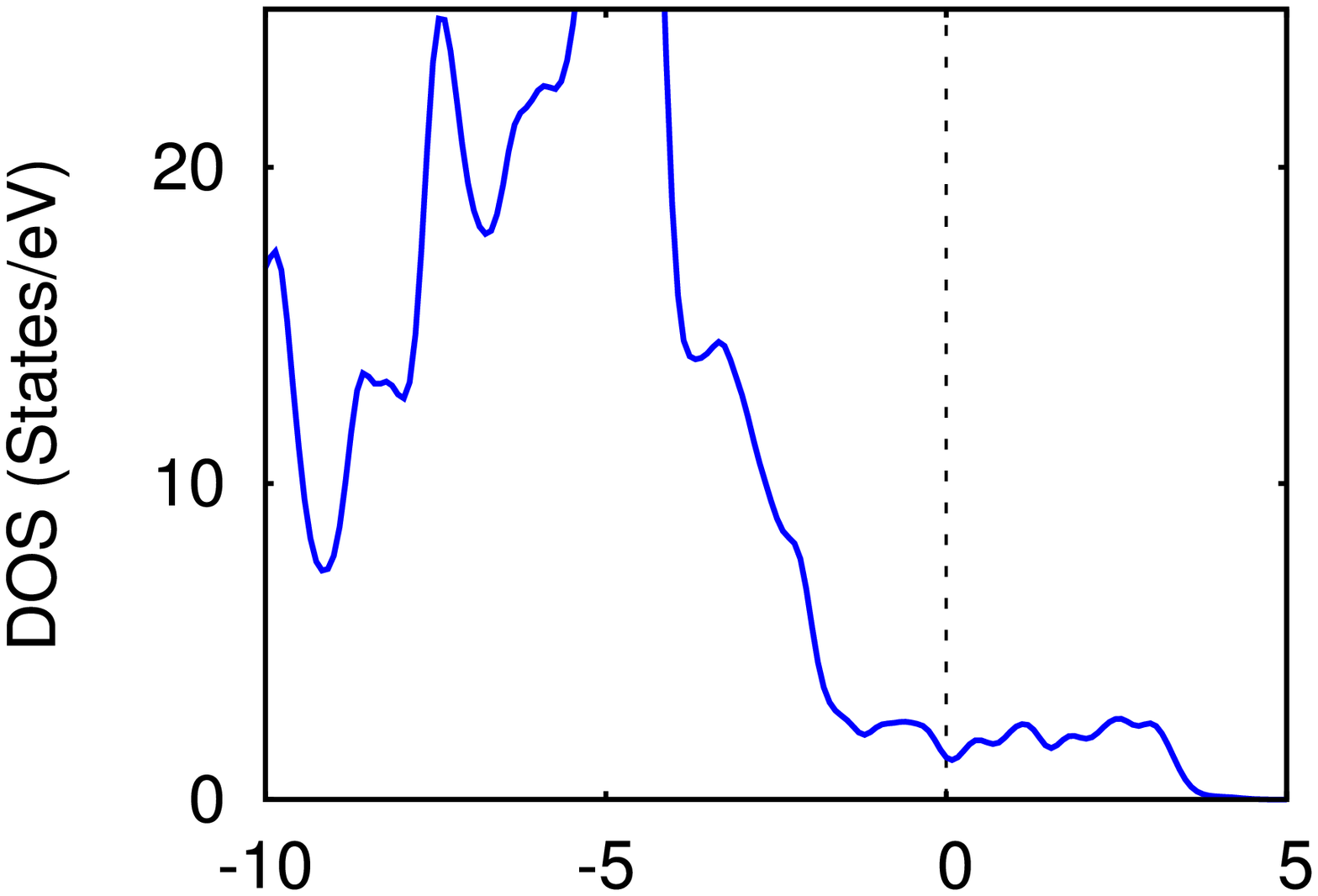}\\
    (g) 60\% hydrogen \hskip 3cm (h) 70\% hydrogen\\~\\
    \includegraphics[width=5.8cm,height=3.5cm]{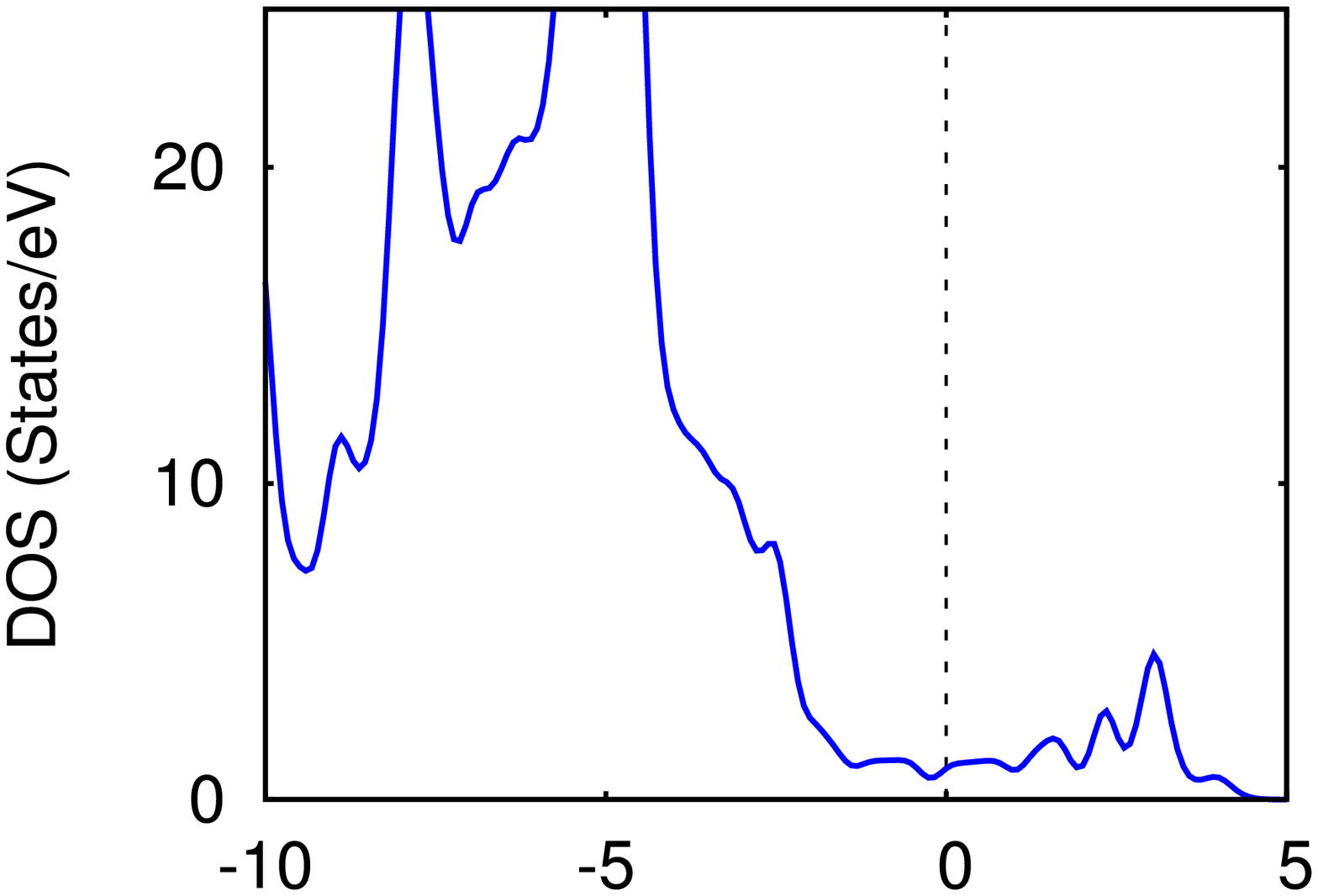}
    \includegraphics[width=5.8cm,height=3.5cm]{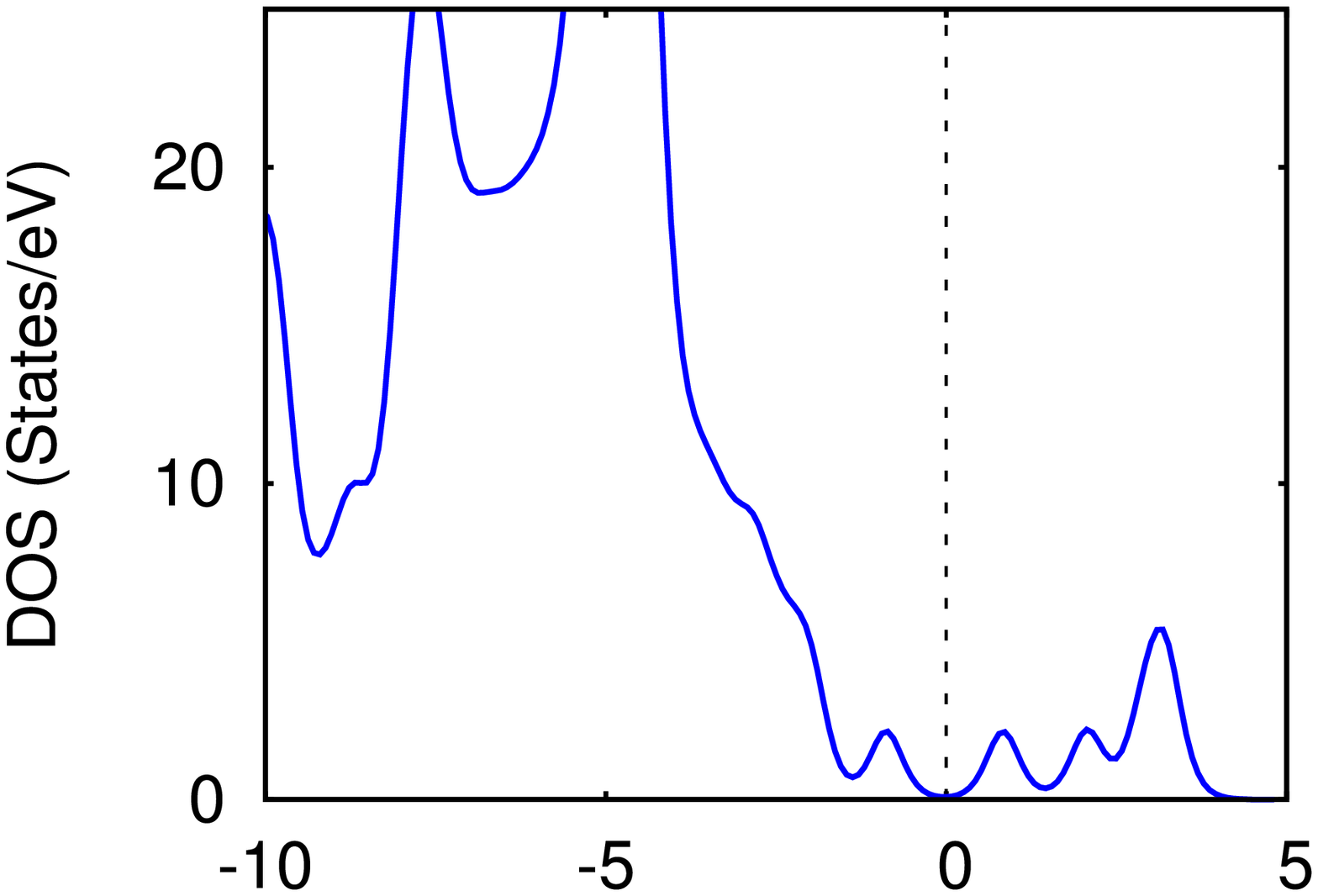}\\
    (i) 80\% hydrogen \hskip 3cm (j) 85\% hydrogen\\~\\
    \vskip 0cm
    \end{center}

\caption{\label{fig2} (Colour online) The total DOS for hydrogenated graphene for various
hydrogen concentrations below 90\%. The zero of the energy is taken at the Fermi level and
is marked by a vertical line. X axis denotes $E-E_f$.}

\end{figure*}

\begin{figure*}
  \begin{center}
    \includegraphics[width=5.6cm,height=3.5cm]{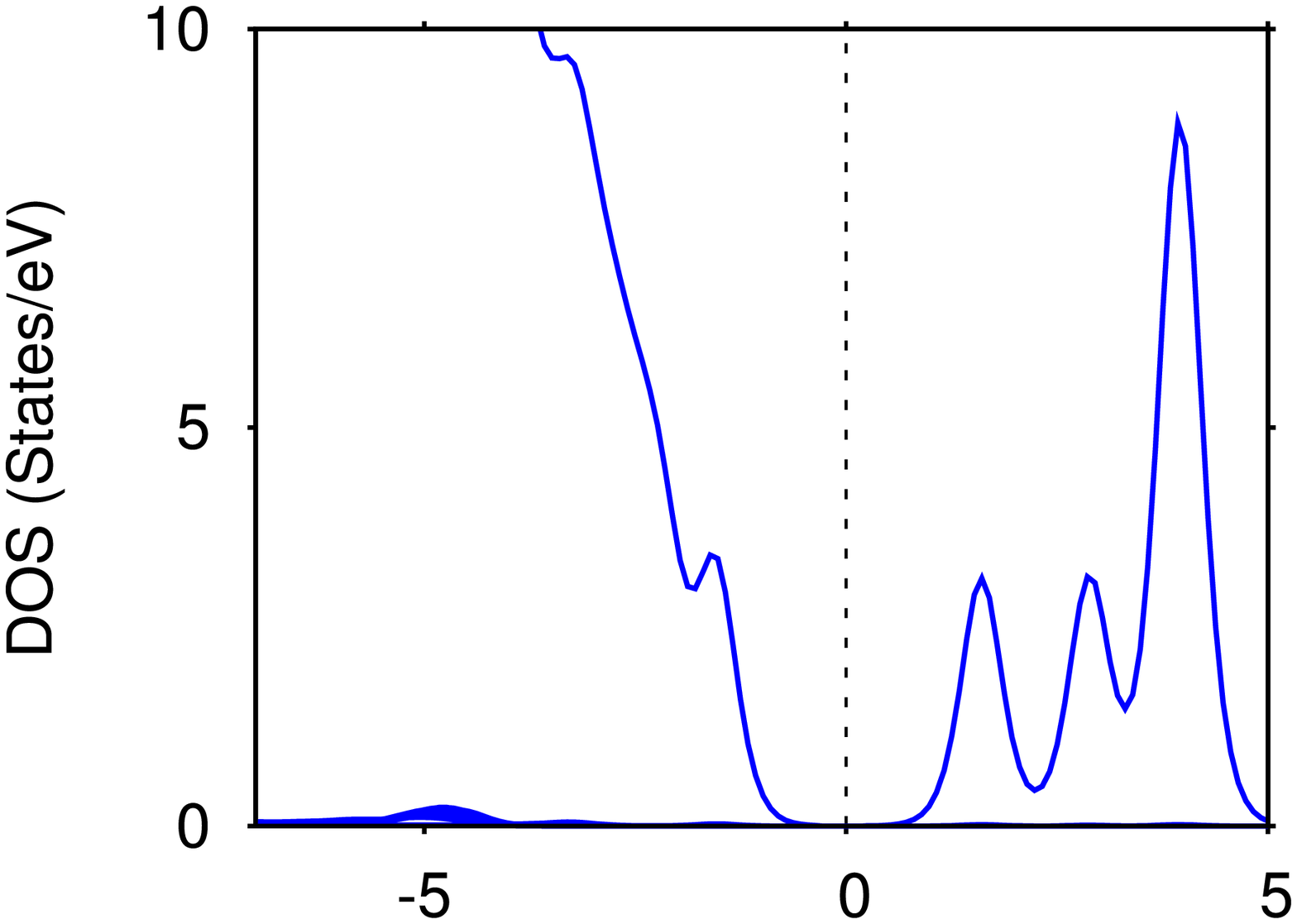}
    \includegraphics[width=5.6cm,height=3.5cm]{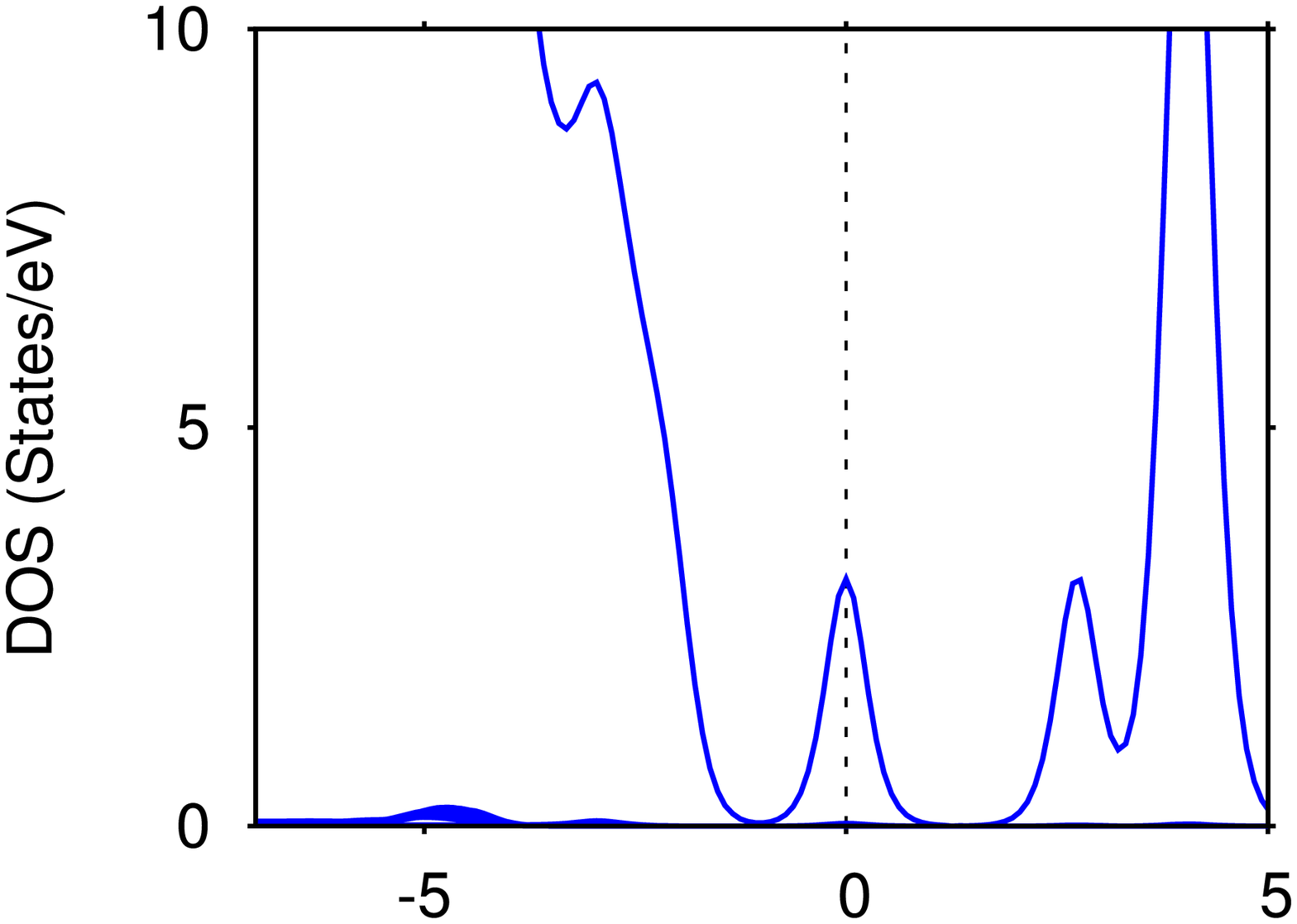}\\
     (a) 92\% hydrogen \hskip 3cm (b) 94\% hydrogen\\~\\
    \includegraphics[width=5.6cm,height=3.5cm]{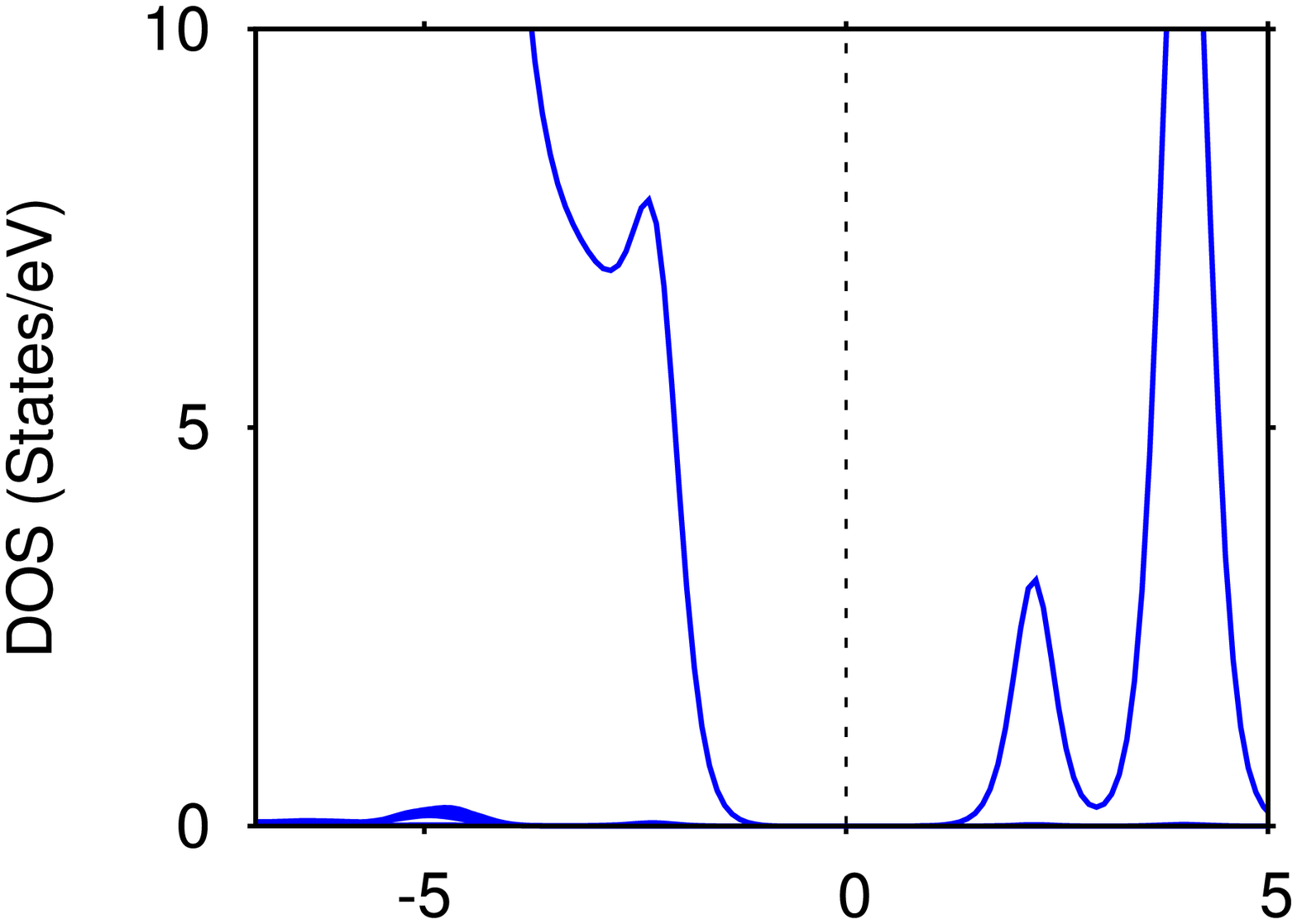}
    \includegraphics[width=5.6cm,height=3.5cm]{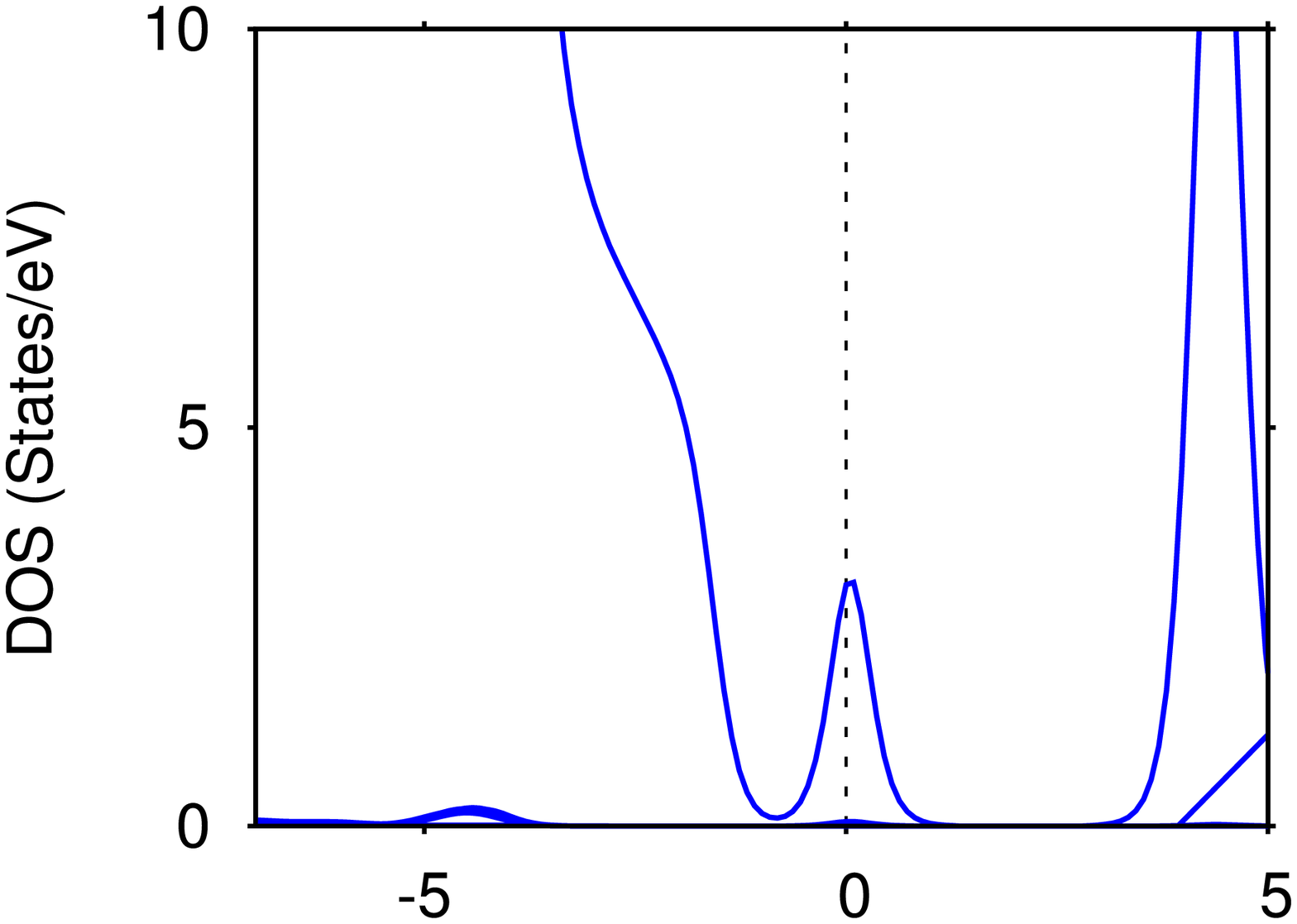}\\
    (c) 96\% hydrogen \hskip 3cm (d) 98\% hydrogen\\~\\
    \includegraphics[width=5.6cm,height=3.5cm]{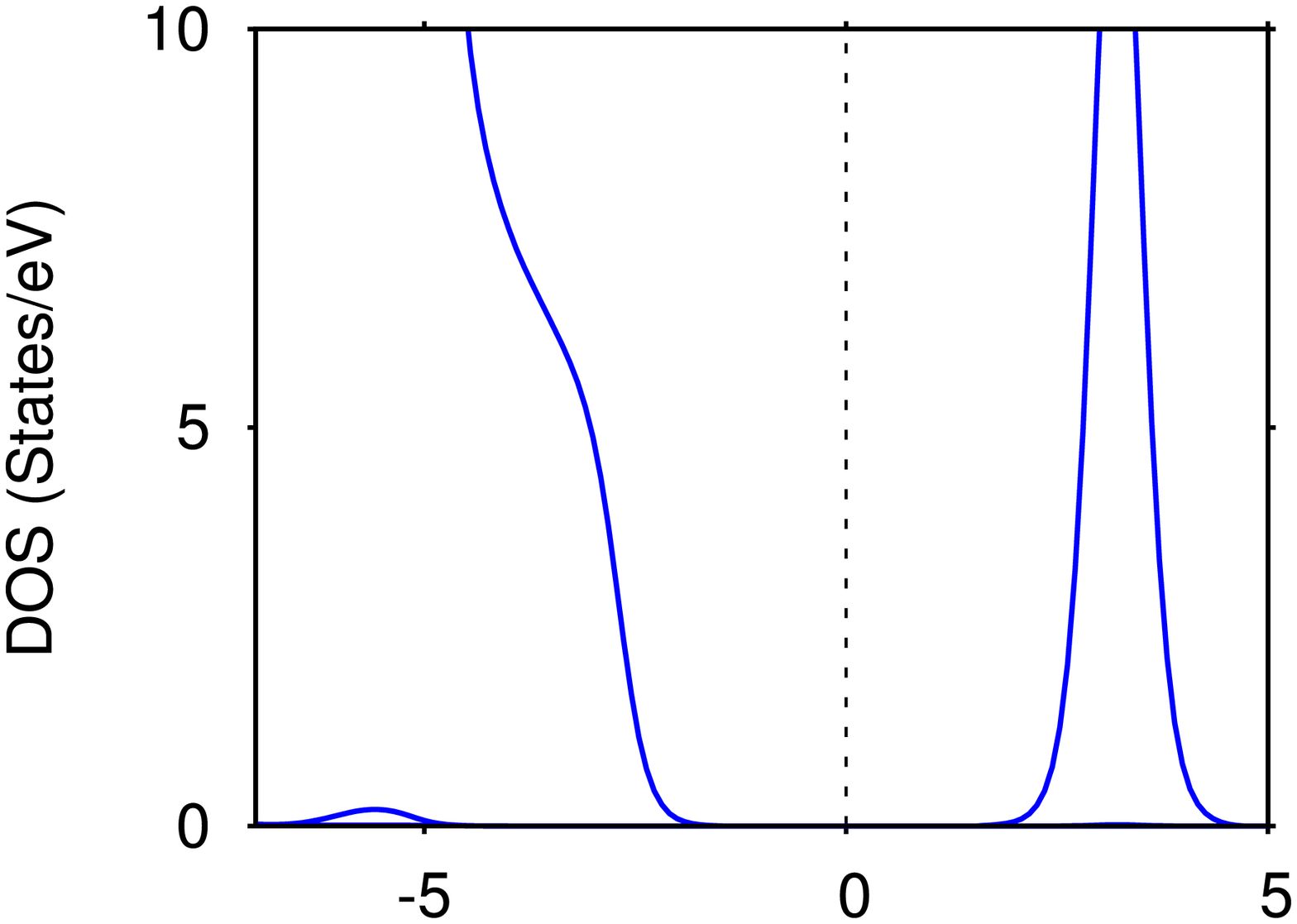}\\
    (e) \hskip 0cm{\it Graphane}
     
  \end{center} 

\caption{\label{fig3} (Colour online) The total DOS for hydrogenated graphene for various
hydrogen concentrations above 90\%. The zero of the energy is taken at the Fermi level and
is marked by a vertical line. X axis denotes $E-E_f$.}

\end{figure*}

Now we discuss the total DOS for different hydrogen coverages which are shown in figure
\ref{fig2} and \ref{fig3}. All the DOS are for the non spin polarised cases. The geometry
used is for the minimum energy configuration (compact cluster of hydrogen atoms). In both
the figures the plots are shown in a restricted region to enhance the clarity near the
Fermi energy.  The effect of addition of a small concentration of hydrogen in the V-shape
DOS near E$_f$ can be seen from figure (\ref{fig2}(b) to \ref{fig2}(d)). It can be noted
that the characteristic V shape valley seen in figure \ref{fig2}(a) is due to peculiar
sp$^2$ bonded  carbon atoms in graphene, and the addition of hydrogen atoms immediately
distorts the symmetry producing the deformation in the DOS. Thus in this region, the DOS
is modified by additional localized $\sigma$-p$_z$ bonds between the hydrogen and carbon.

As the hydrogen coverage increases there is a significant increase in the value of DOS at
the Fermi level. The process of hydrogenation is accompanied by the change in the
geometry.  The hydrogenated carbon atoms are now moved out of the graphene plane, in turn
the lattice is distorted and the symmetry is broken. As a consequence, more and more {\bf
$k$} points in the Brillouin zone contribute to the DOS near Fermi level, the increase
being rather sharp after 20\% coverage.  The region ranging from 30\% coverage to about
60\% coverage is characterized by the finite DOS of the order of 2.5/eV near the Fermi
energy. As we shall discuss this region can be described as having metallic character with
delocalized charge density. 

Now we complete the discussion of DOS by presenting the cases of hydrogen coverages
greater than 90\%. In figure \ref{fig3} we show the density of states obtained by the
removal of one - four hydrogen atoms in a unit cell containing fifty carbon atoms.
Clearly one and three hydrogen vacancies (i.e 98\% hydrogen and 94\% hydrogen
respectively) induce states on the Fermi level while two and four hydrogen vacancies (i.e
96\% hydrogen and 92\% hydrogen respectively) do not induce any state at the Fermi level
(See the discussion of hydrogen imbalance later). For a small number of vacancies these
are $\pi$ bonded states localized on bare carbon atoms.

In summary, it is possible to discern, rather broadly, three regions of hydrogen coverage.
A low concentration region, where the DOS undergo the distortions due to loss of lattice
symmetry of the system, intermediate concentration region where a metallic-like phase is
seen and a vary high concentration region where most of the carbon atoms are hydrogenated
and vacancy gives rise to midgap states.

As discussed before, our calculations show the presence of localized states in the DOS for
a very low hydrogen coverage as well as for very high hydrogen coverage (midgap states).
These DOS show a pattern of peak or a valley at E$_f$ (in case of low coverage) and at the
centre of the gap (in case of high coverage). The reasons can be attributed to the
existence of sublattice imbalance. If the difference of hydrogen atoms on the two sides of
the sheet is odd then it leads to a peak. Hence by adding impurities one by one to {\it
graphane} we get a sequence of midgap states. This is in consistent with the work of
Casolo \etal~\cite{casolo}. Interestingly this feature is also retained  for the
intermediate coverages of hydrogen (where we get a finite DOS at the Fermi level) when the
sub lattice imbalance of hydrogen is one, e.g., in figure \ref{fig2}(e), \ref{fig2}(f) and
\ref{fig2}(g), we see the peaks at Fermi level.  As we shall see in section \ref{sec:mag}
this feature has an implication for the magnetic behaviour of the system.

\begin{figure} 
\centerline{\includegraphics[width=8cm]{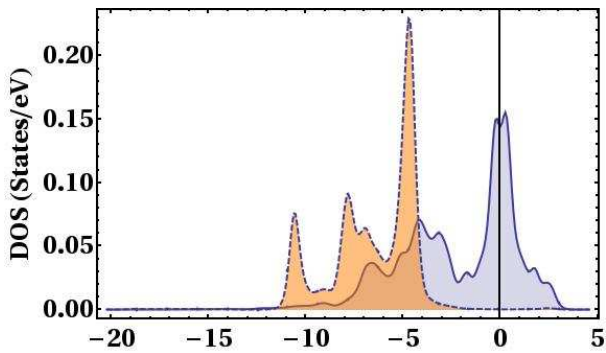}  }

\caption{\label{fig4-pdos}(Colour online) Site projected DOS for hydrogenated carbon sites
(dotted line) and bare carbon sites (continuous line) for hydrogen coverage of 40\%.
Almost all the contribution comes from the $p_z$ which has been showed in the figure. Note
that only bare carbon atoms contribute to the DOS at Fermi level.}

\end{figure}

In order to bring out the difference between the local electronic structure of the bare
carbon atoms and the hydrogenated carbon atoms, we have  analyzed the site projected DOS
for all the cases. In figure \ref{fig4-pdos}, we show site projected DOS for 40\% hydrogen
coverage  depicting the contributions from a hydrogenated carbon site and a bare carbon
site. Quite clearly, the contribution around the Fermi level comes from the $p_z$ orbitals
of bare carbon atoms only.  This is a general feature for all the systems investigated.
It may be emphasized  that in pure graphene all the carbon atoms contribute to a single
{\bf $k$} point (Dirac point).  In contrast to this, upon hydrogenation, only bare carbon
atoms contribute to the DOS at Fermi energy, and they do so at many {\bf $k$} points of
the Brillouin zone.  As the concentration increases further (above 80\%), there are too
few bare carbon atoms available for the formation of delocalized $\pi$ bonds. The value of
DOS approaches zero and a gap is established with a few mid gap states.

\begin{figure}
  \begin{center}
    \includegraphics[width=8cm]{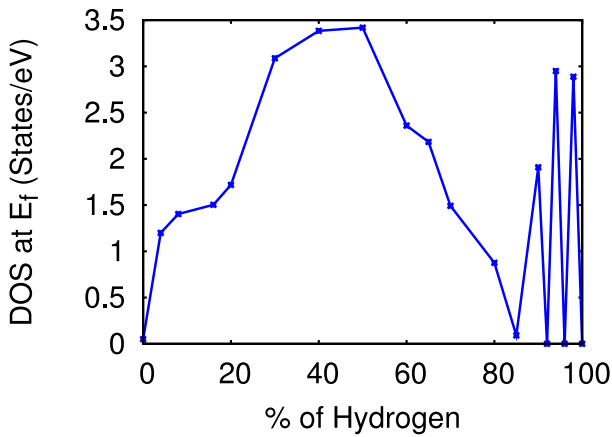}
  \end{center}

\caption{\label{fig5} (Colour online) Variation of the value of DOS at the Fermi level as
a function of hydrogen coverage.}

\end{figure}

The evolution towards the metallic state can be better appreciated by examining the
variation in the value of DOS at the Fermi level which is shown in figure \ref{fig5}.  A
clear rise in DOS is seen after 20\% hydrogen concentration peaking around 3.5 eV at 50\%
concentration. This rise is due to the increasing number of $k$ points contributing to the
Fermi level, as inferred from the analysis of the individual bands.  Evidently, over a
significant range of concentration, the value at Fermi level is more than 2/eV. The
decline seen after 60\% is because of the reduction in the number of bare carbon atoms.
The character of the DOS changes after about 80\%. The value of the DOS at the Fermi level
oscillates between zero and some finite value. It is most convenient to describe this
region as {\it graphane} with defects by the removal of a few hydrogen atoms giving rise
to mid gap states. The nature and the placement of the induced states is dependent on the
number of hydrogen atoms removed. 

\begin{figure*}
  \begin{center}
    \includegraphics[width=5.5cm]{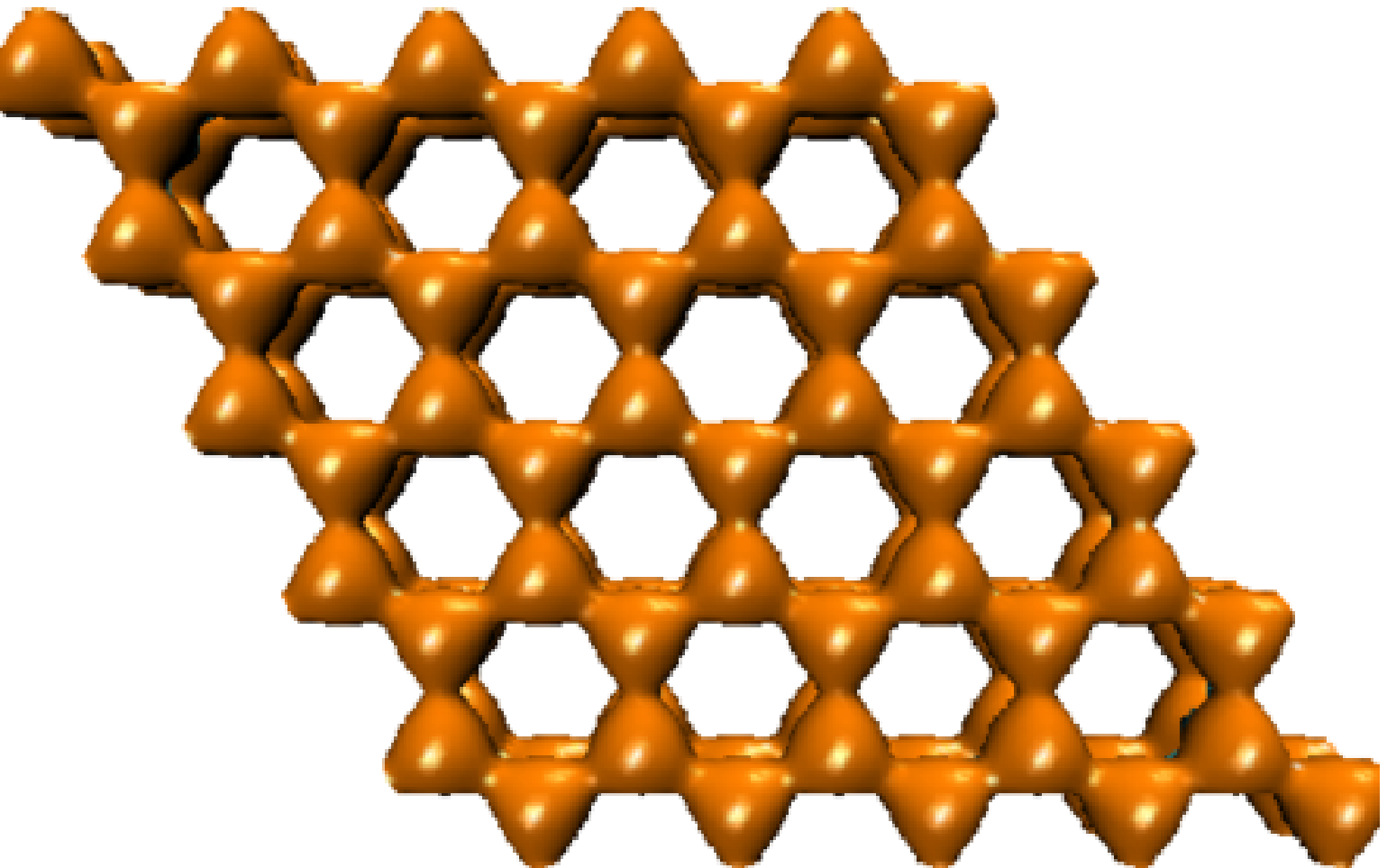}
    \includegraphics[width=5.5cm]{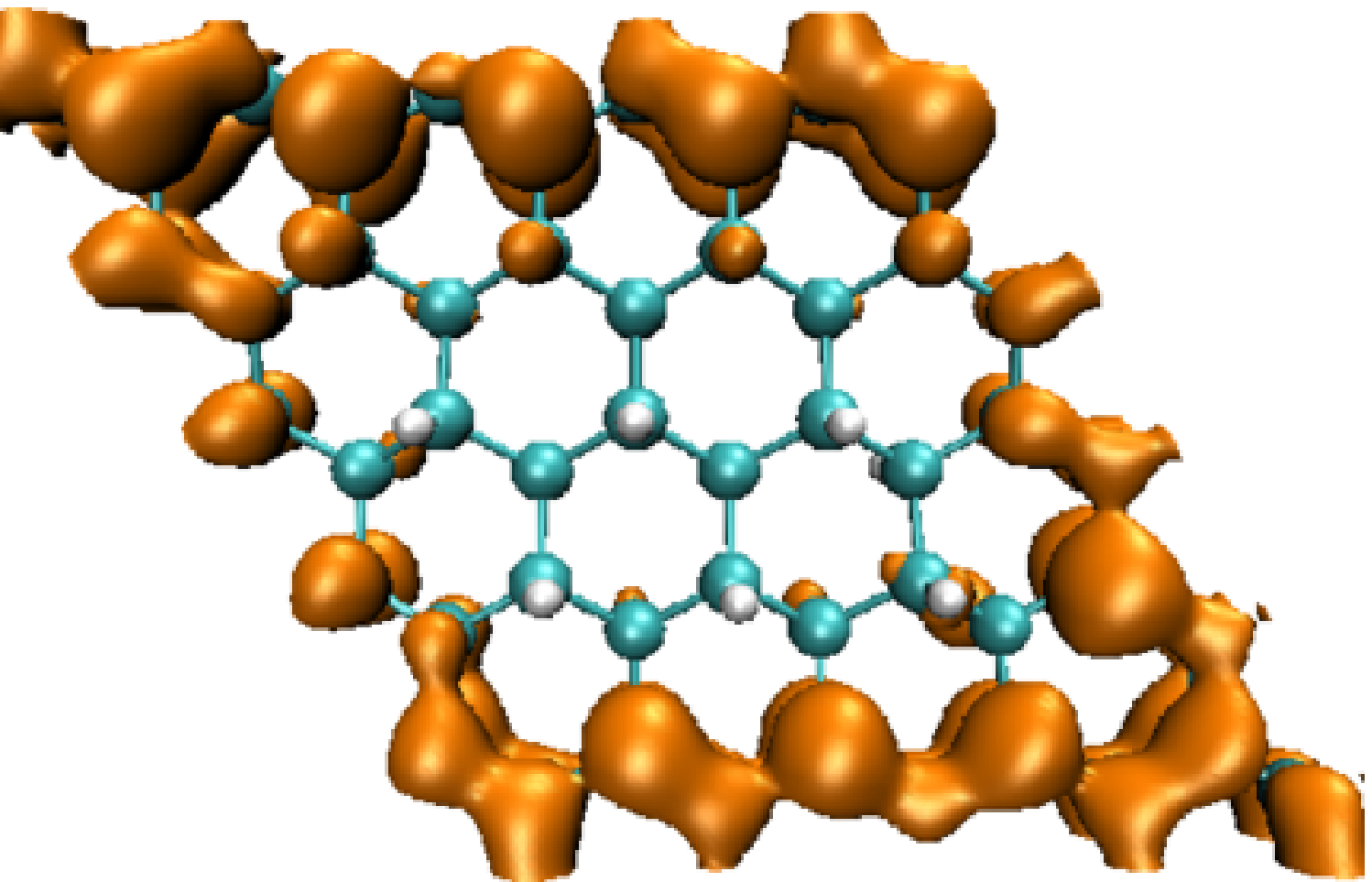} \\
    (a) Graphene \hskip 3cm (b) 40\%  hydrogen\\~\\
    \includegraphics[width=5.5cm]{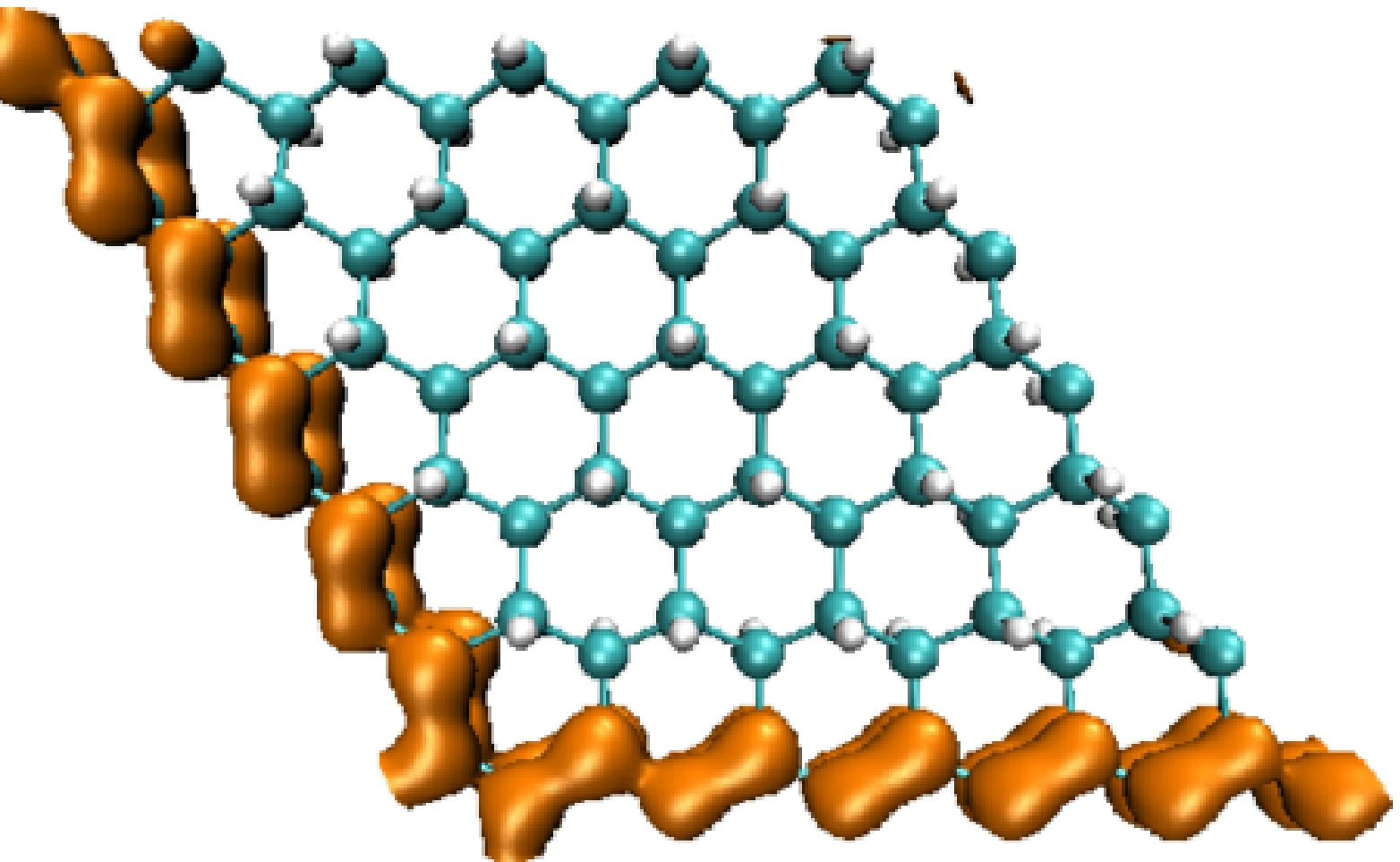}
    \includegraphics[width=5.5cm]{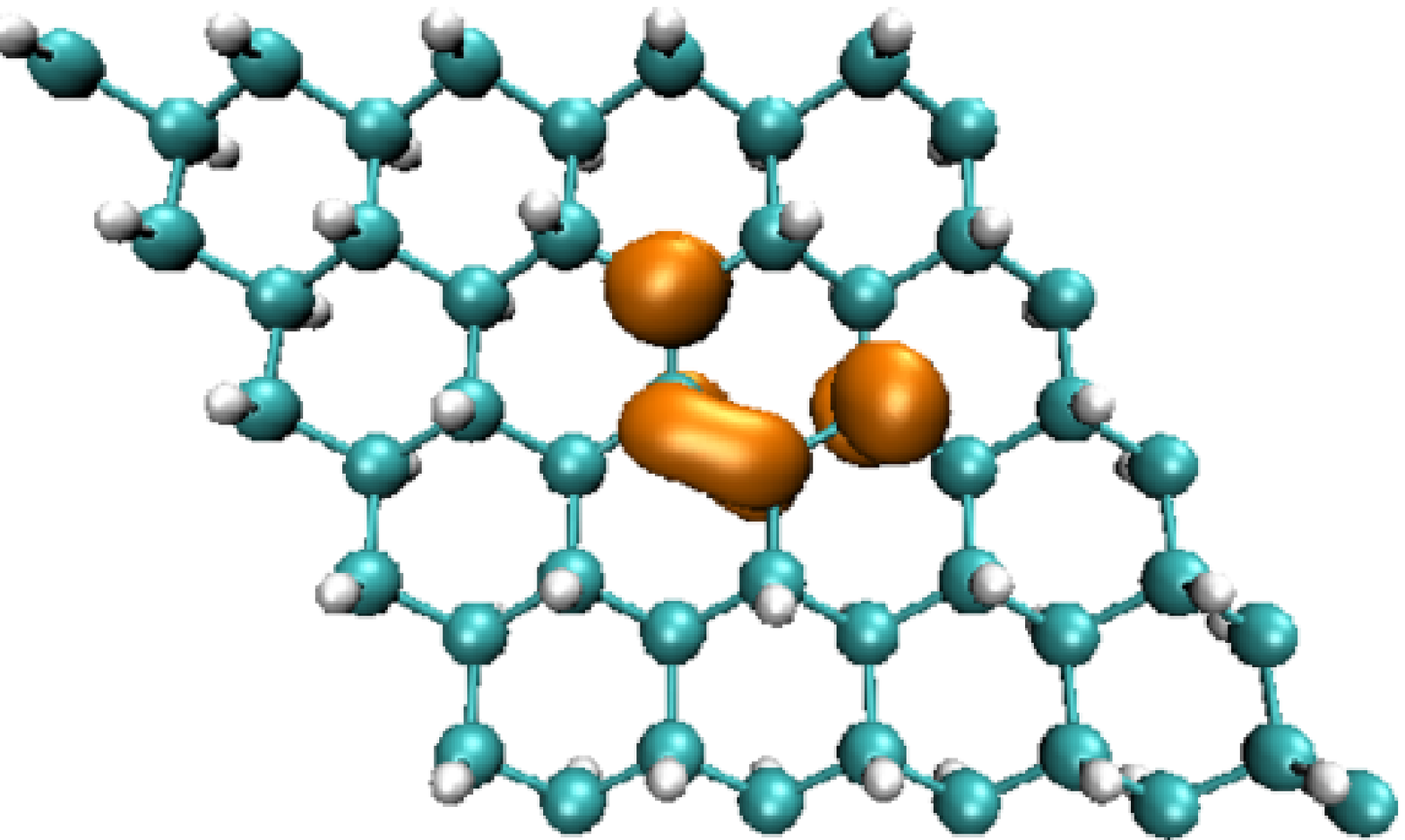}\\
    (c) 70\% hydrogen \hskip 3cm (d) 92\%  hydrogen\\~\\
    \end{center}
\caption{\label{fig6} (Colour online) Isosurfaces of charge densities of bands near the
Fermi level (see text). For comparison, the charge density of graphene at Dirac point is
also shown.}
\end{figure*}

It may be emphasized  that the presence of states around the Fermi level giving finite DOS
does not guarantee that the system is metallic unless we examine the nature of
localization of the individual states. Therefore we have examined the energy resolved
charge densities of the states near the Fermi level and the electron localization function
(ELF). The energy resolved charge densities are obtained by summing the charge densities
of all the {\bf k} points contributing in the small energy region near E$_f$. Therefore,
the charge densities shown in (figure \ref{fig6}) for 40\%, 70\% and 92\% hydrogen
coverages specifically brings out the nature of the states near E$_f$. A particularly
striking feature is the formation of two spatially separated regions as seen in figure
\ref{fig6}(b) and \ref{fig6}(c).  The hydrogenated regions hardly contribute to the charge
density giving rise to the insulating regions surrounded by the $\pi$ bonded bare carbon
atoms forming conducting regions.  This feature is also seen for the higher concentrations
upto 70\%.  It may be emphasized that the topology in this range of concentrations
(30\%-70\%) shares a common feature namely, there is a contiguous region formed by the
bare carbon atoms.  It may be pointed out that the contiguous charge density is attributed
to the favoured configuration of compact cluster formation \footnote{It is our conjecture,
based on results of classical bond percolation on 2D hexagonal lattice, that below
hydrogen concentration of 61.5\% (randomly distributed), the bare carbon atoms will form
continuous chains. However, we have not carried out any calculations to verify this.}.
The change in the character of the state at 90\% and above is also evident in figure
\ref{fig6}(d).  There are insufficient number of bare carbon atoms to form  contiguous
regions.  As a consequence, these carbon atoms form localized bonds giving rise to mid gap
states noted earlier. 

\begin{figure*}
\begin{center}
\includegraphics[width=6.0cm,height=4.5cm]{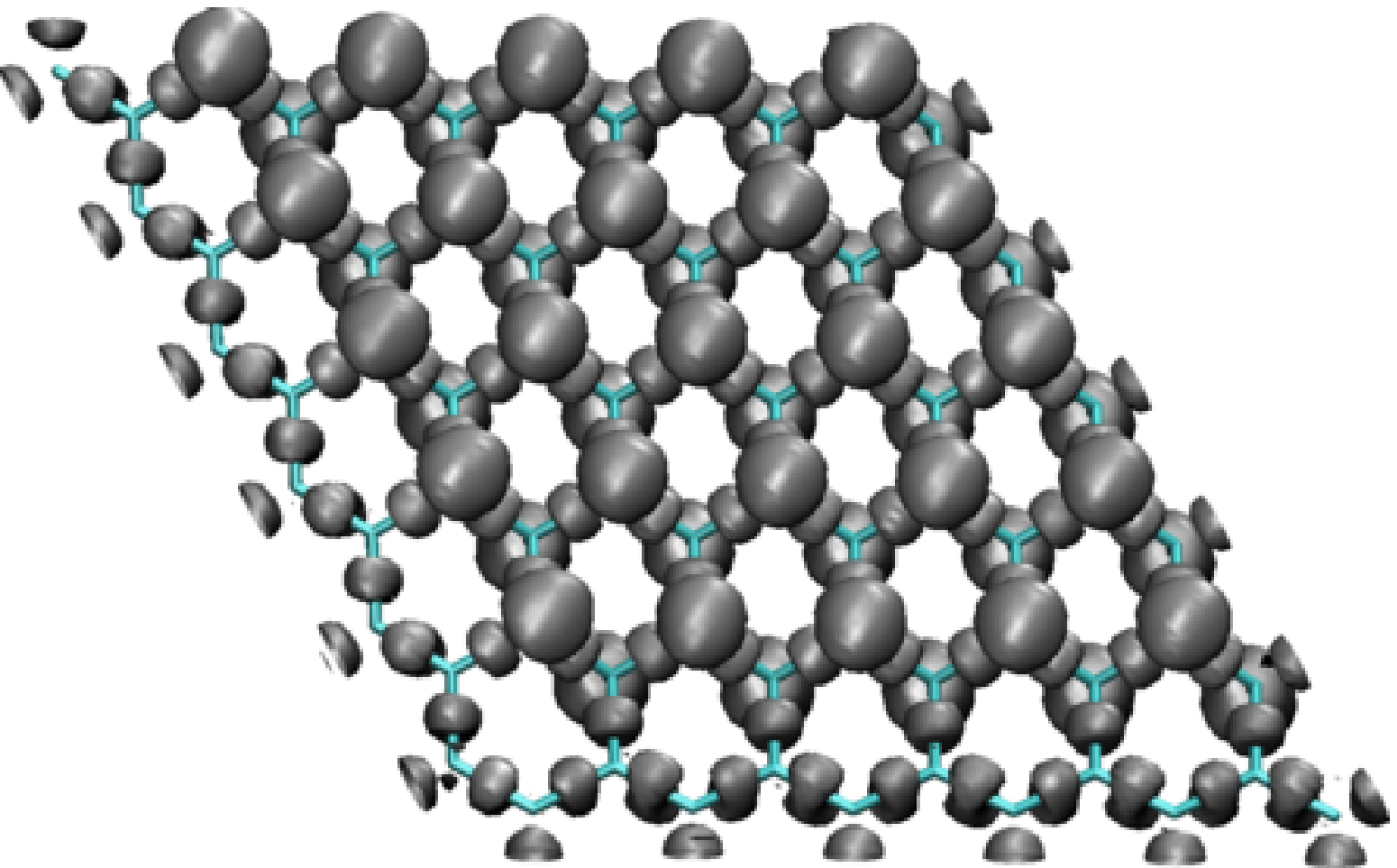}
\includegraphics[width=6.5cm,height=4.5cm]{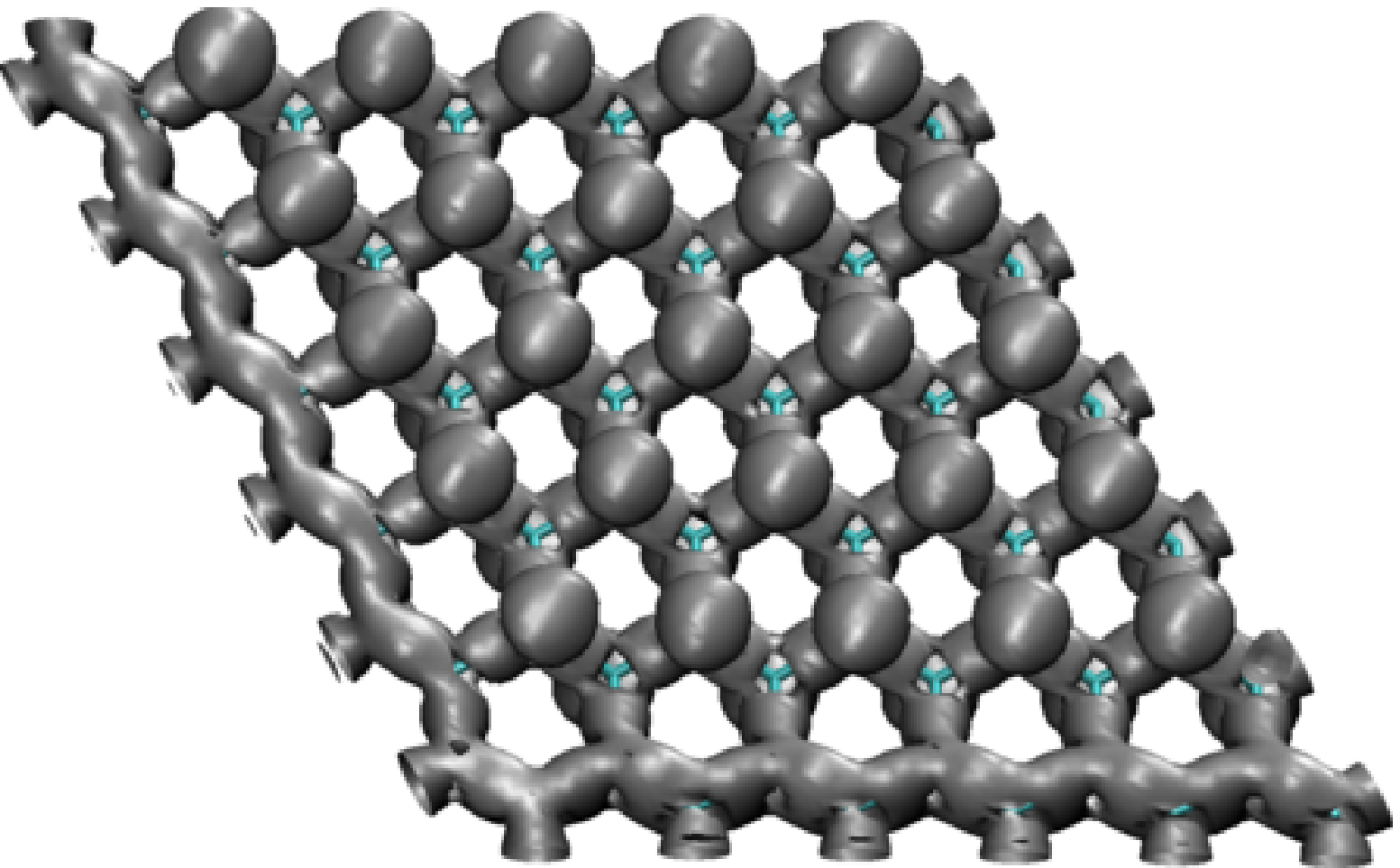} \\
(a) \hskip 5cm (b) \\~\\
\end{center}

\caption{\label{fig7} (Colour online) ELF plots with two different isosurface values for
70\% hydrogen concentration. Figure (a) shows for isosurface value 0.75 and figure (b)
shows for isosurface value 0.62 .}

\end{figure*}

The degree of delocalization of an electron of the system can be understood by examining
the electron localization function as follows.  For a single determinantal wave function
built from Kohn-Sham orbitals, $\psi _{i}$, the ELF is defined as,~\cite{becke}

\begin{equation} \chi _{{\rm ELF}}=[1+{(D/D}_{h}{)}^{2}]^{-1}, \end{equation} where
\begin{eqnarray} D_{h}&=&(3/10){(3{\pi }^{2})}^{5/3}{\rho }^{5/3}, \\ D&=&(1/2)\sum_{i}{\
|{\bm{\nabla} \psi _{i}}|}^{2}-(1/8){|{\bm{\nabla} \rho }|}^{2}/\rho, \end{eqnarray} 

with $\rho \equiv \rho ({\bf r})$ being the electron density.  D is the excess local
kinetic energy density due to Pauli repulsion and D$_{h}$ is the Thomas--Fermi kinetic
energy density. The numerical values of $\chi _{{\rm ELF}}$ are conveniently normalized to
a value between zero and unity. A value of 1 represents a perfect localization of the
charge, while the value for the uniform electron gas is 0.5.  Typically, the existence of
an isosurface of a high value of $\chi _{{\rm ELF}}$ say, $\ge 0.70$, signifies a
localized bond. 

We have examined the isosurfaces of ELF for various values between 0.5 and 1.0. The
isosurface for a typical value 0.75, is shown in figure \ref{fig7}(a) for 70\% coverage.
It can be seen that for this high value isosurface exists mainly along carbon-hydrogen
bonds ($\sigma$-p$_z$). The existence of localized $\sigma$ bonds between the bare carbon
atoms can also be noticed. In figure \ref{fig7}(b) we show the isosurface for the value of
0.62. The figure shows a continuous surface covering all the bare carbon atoms. This
signifies the delocalized nature of the charge density arising out of $\pi$ bonds formed
by p$_z$ orbitals.  We have analyzed the ELF for all the coverages. The above features are
seen for the coverages from 30\% to 80\%.  Thus the analysis of ELF confirms the
delocalized nature of the charge density near the Fermi level for the concentration from
30\% to 80\%.

\subsection{\label{sec:mag}Magnetism}

\begin{figure*}
\begin{center}
\includegraphics[width=6.0cm,height=4.5cm]{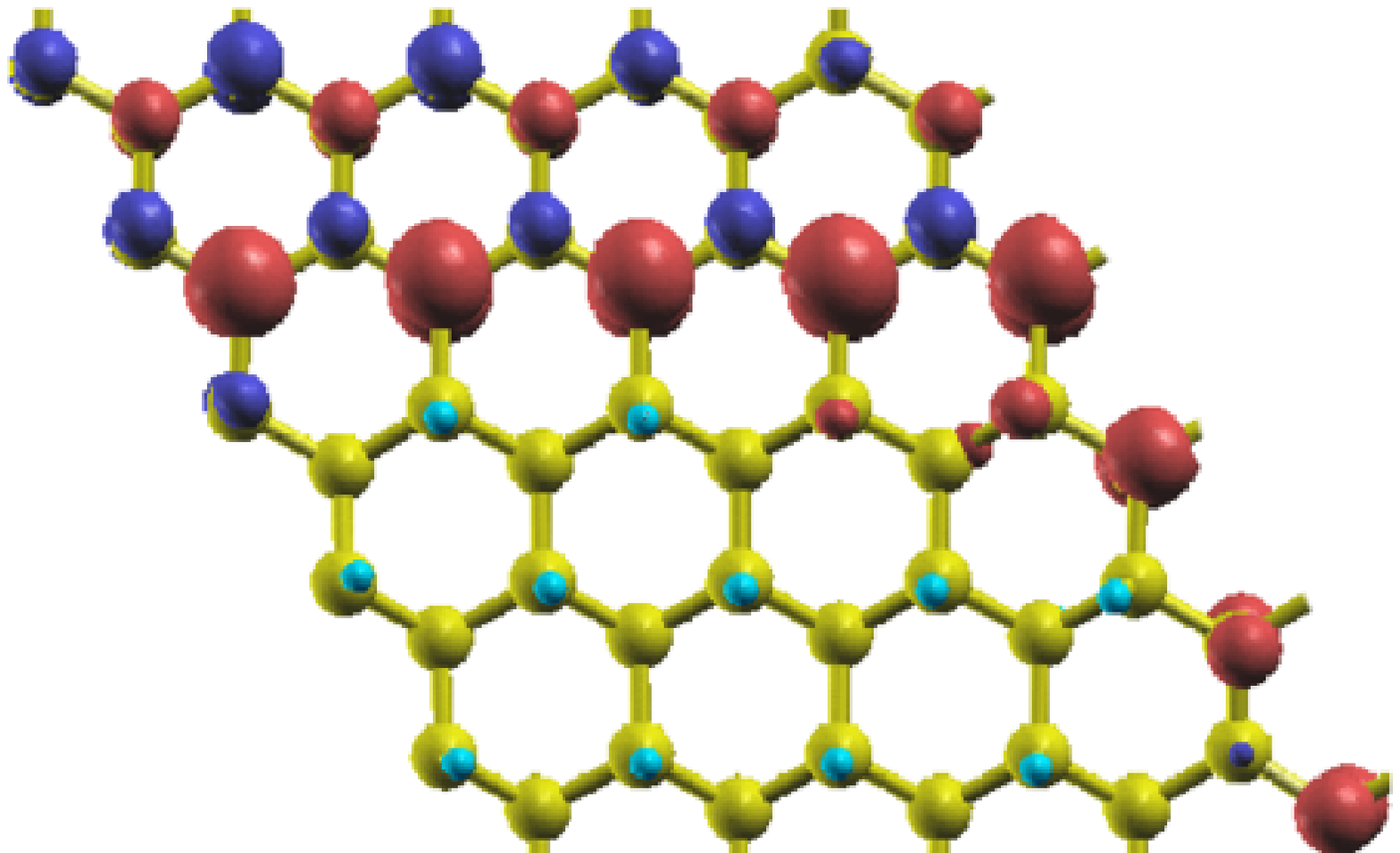}
\includegraphics[width=6.5cm,height=4.5cm]{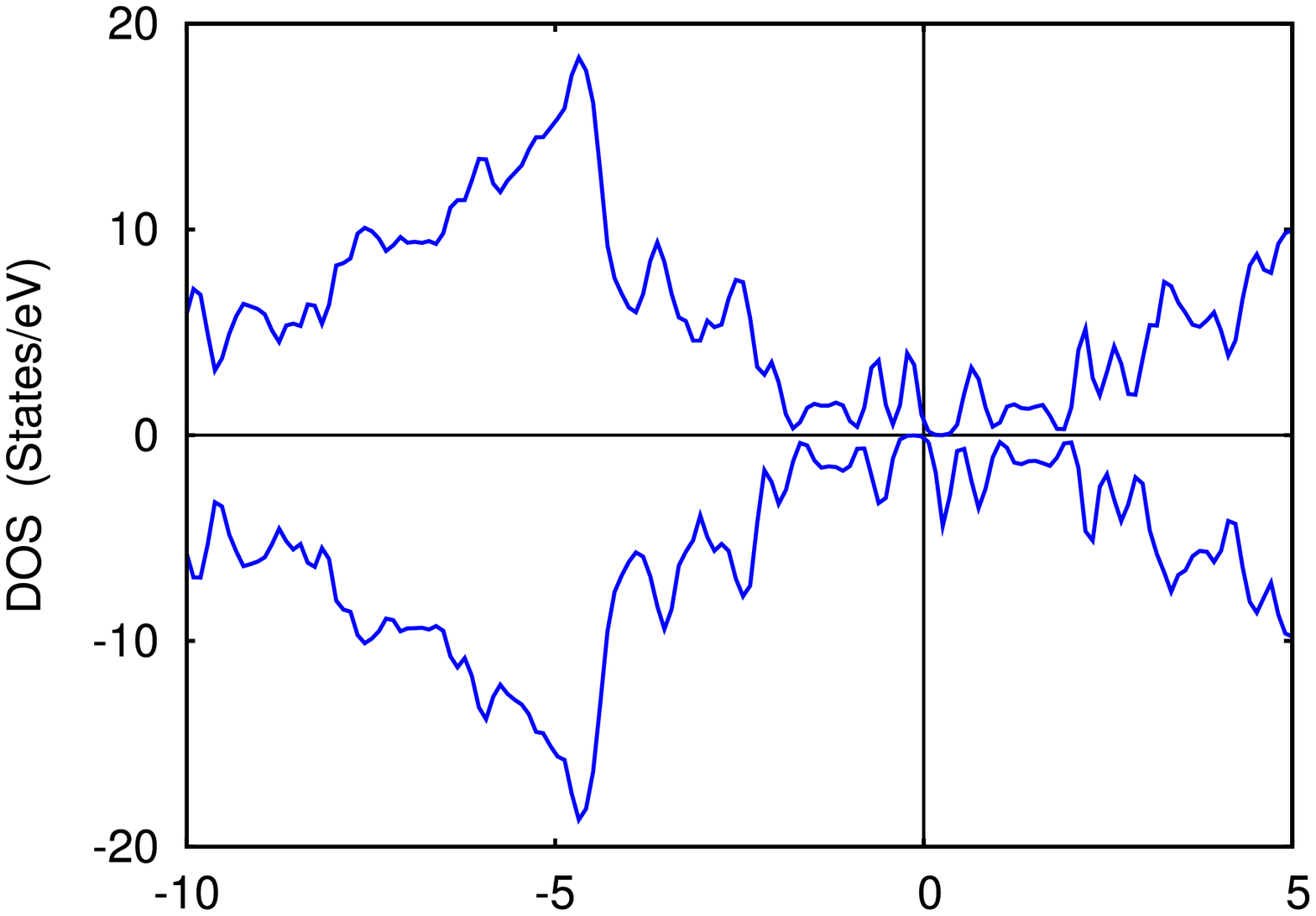} \\
(a) \hskip 5cm (b) \\~\\
\includegraphics[width=6.5cm,height=4.5cm]{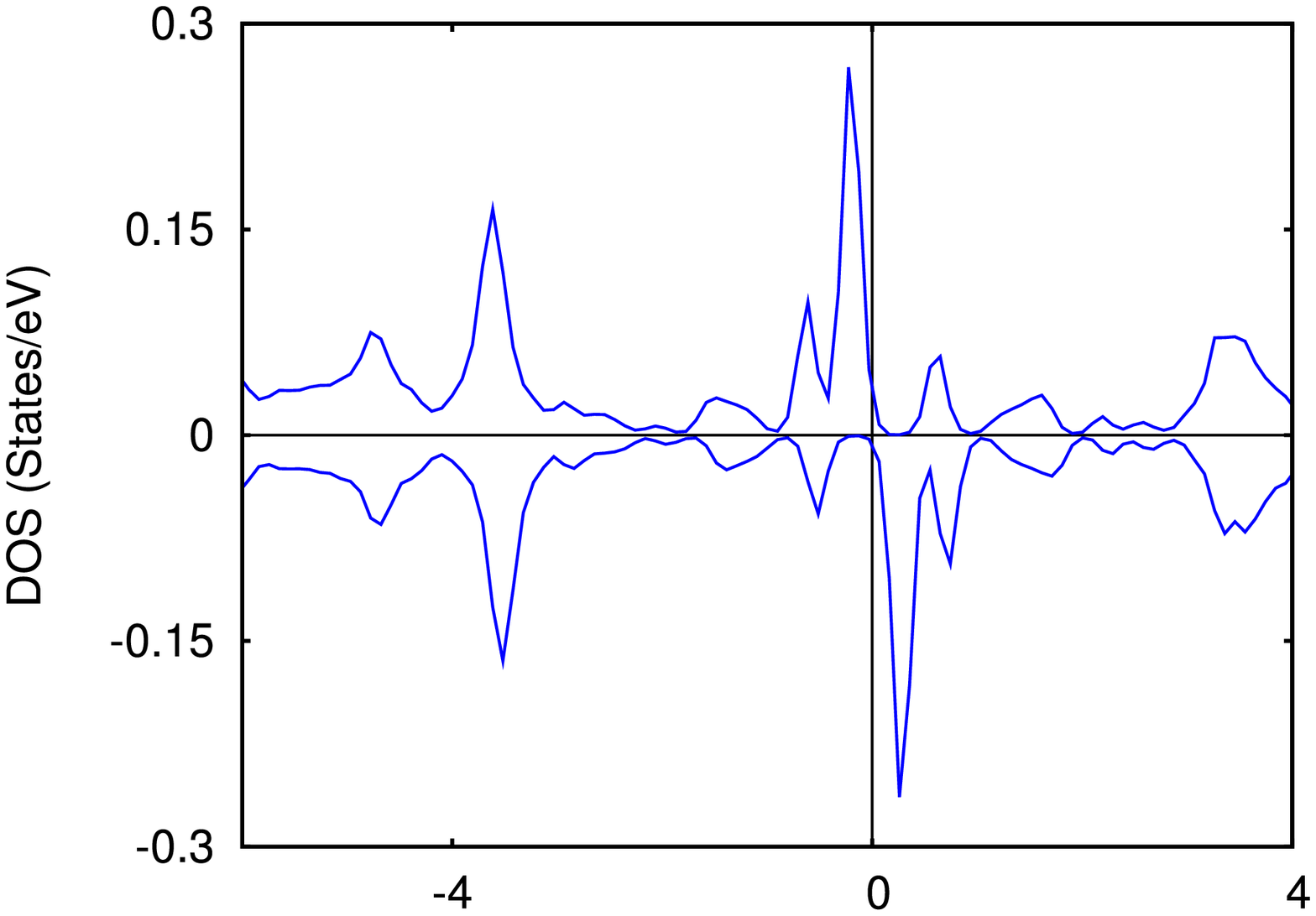} 
\includegraphics[width=6.5cm,height=4.5cm]{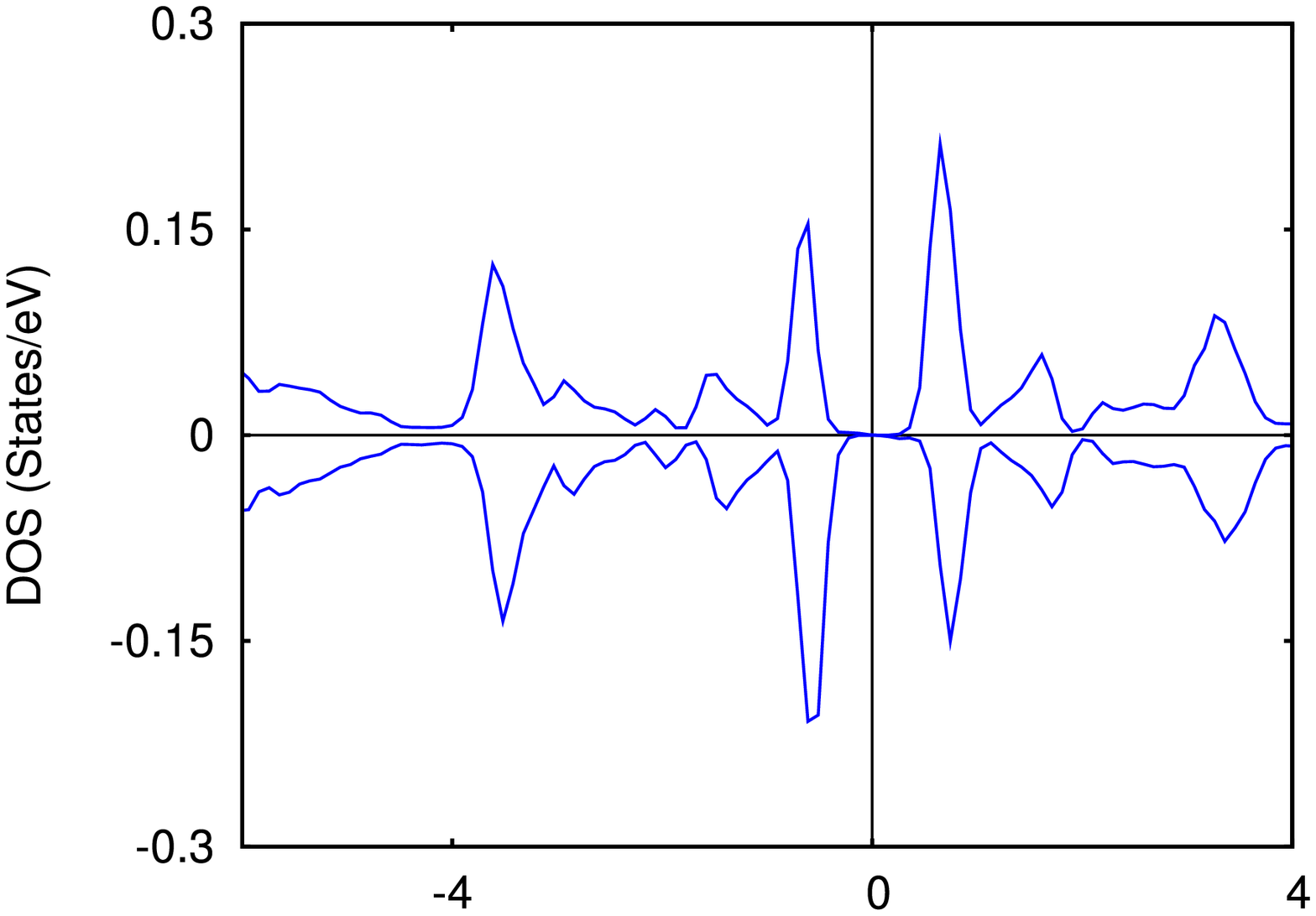} \\
(c) \hskip 5cm (d) \\~\\
\end{center}

\caption{\label{fig8} (Colour online) (a) Spin density, (b) total DOS, (c) and (d) site
projected ($p_z$ only) DOS for two bare carbon atoms from two different sublattices. X
axis of the plots denotes $E-E_f$. This figure is for 50\% hydrogen coverage.}

\end{figure*}

The existence of a peak at Fermi level in the non spin polarized calculation is an
indication of Stoner instability that may lead to a more stable spin polarized solution.
As an example we have analyzed the case of 50\% hydrogen coverage which shows a peak in
the DOS at the Fermi level.  When we allow spin polarization in our calculations, indeed
we find a stable ferromagnetic solution, e.g., for 50\% hydrogen coverage we get a total
magnetic moment $\sim$ 1 $\mu_B$ per unit cell as reflected in the spin density plot shown
in figure \ref{fig8}(a). However, the exchange splitting seen on the bare carbon atoms is
rather small and is expected to survive only at a very low temperature.  This fragile
nature of magnetism is clearly indicated by the collapse of magnetic moment with a
smearing width of 0.08 eV.  The site projected DOS for two bare carbon atoms belonging to
two different sublattices along with the spin density plots are shown in figure
(\ref{fig8}(c) and \ref{fig8}(d)).  Carbon atom belonging to one sublattice shows a
spin-polarized behaviour whereas the other sublattice carbon atom has an energy gap in the
electronic spectrum.  This sublattice effect is similar to what is observed in case of
graphene nanoribbons~\cite{bhandary}. 

\subsection{\label{subsec:tune}Tuning the electronic structure with hydrogenation}

\begin{figure*}
\begin{center}
\includegraphics[width=6.0cm]{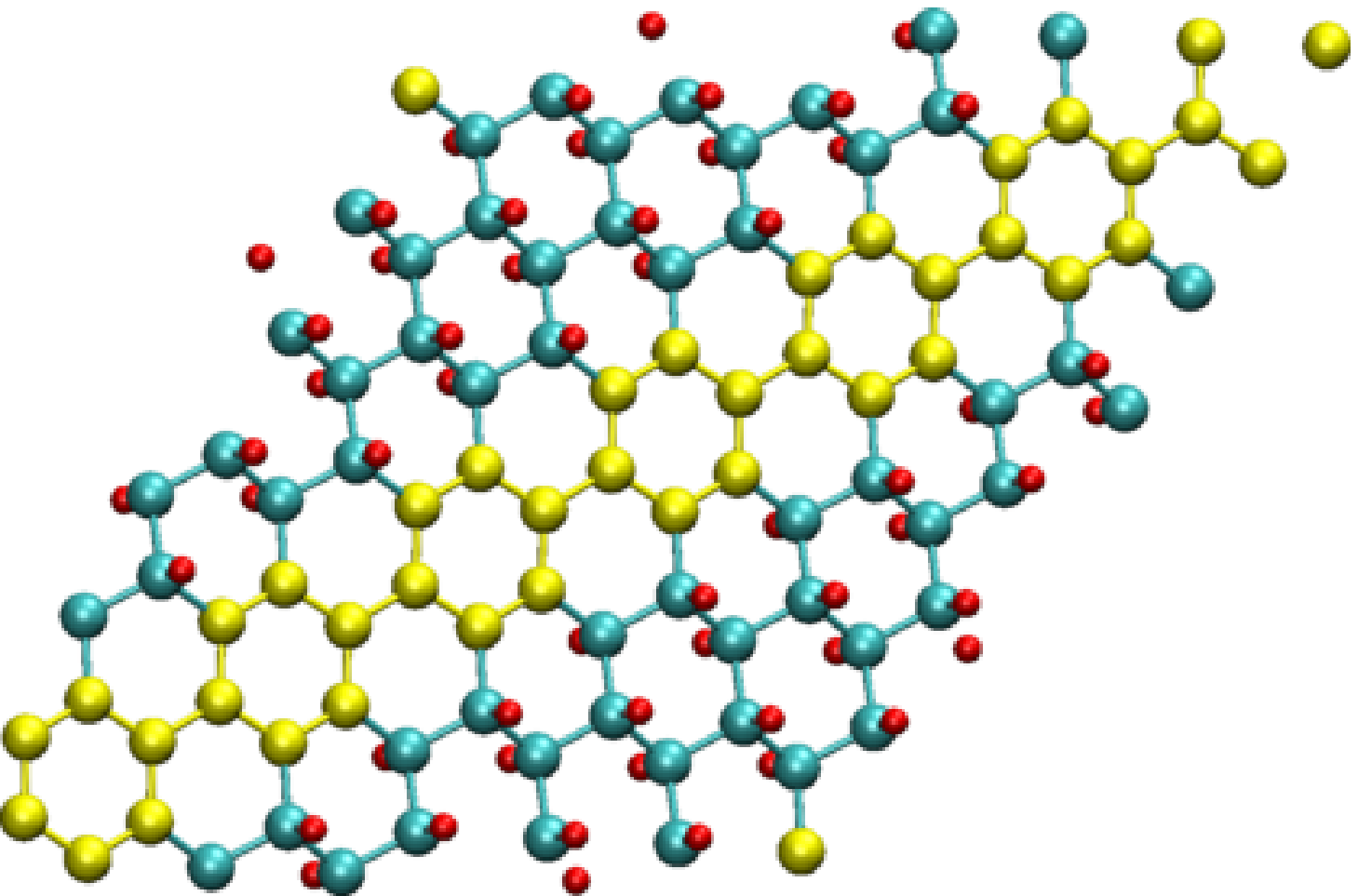}
\includegraphics[width=6.0cm]{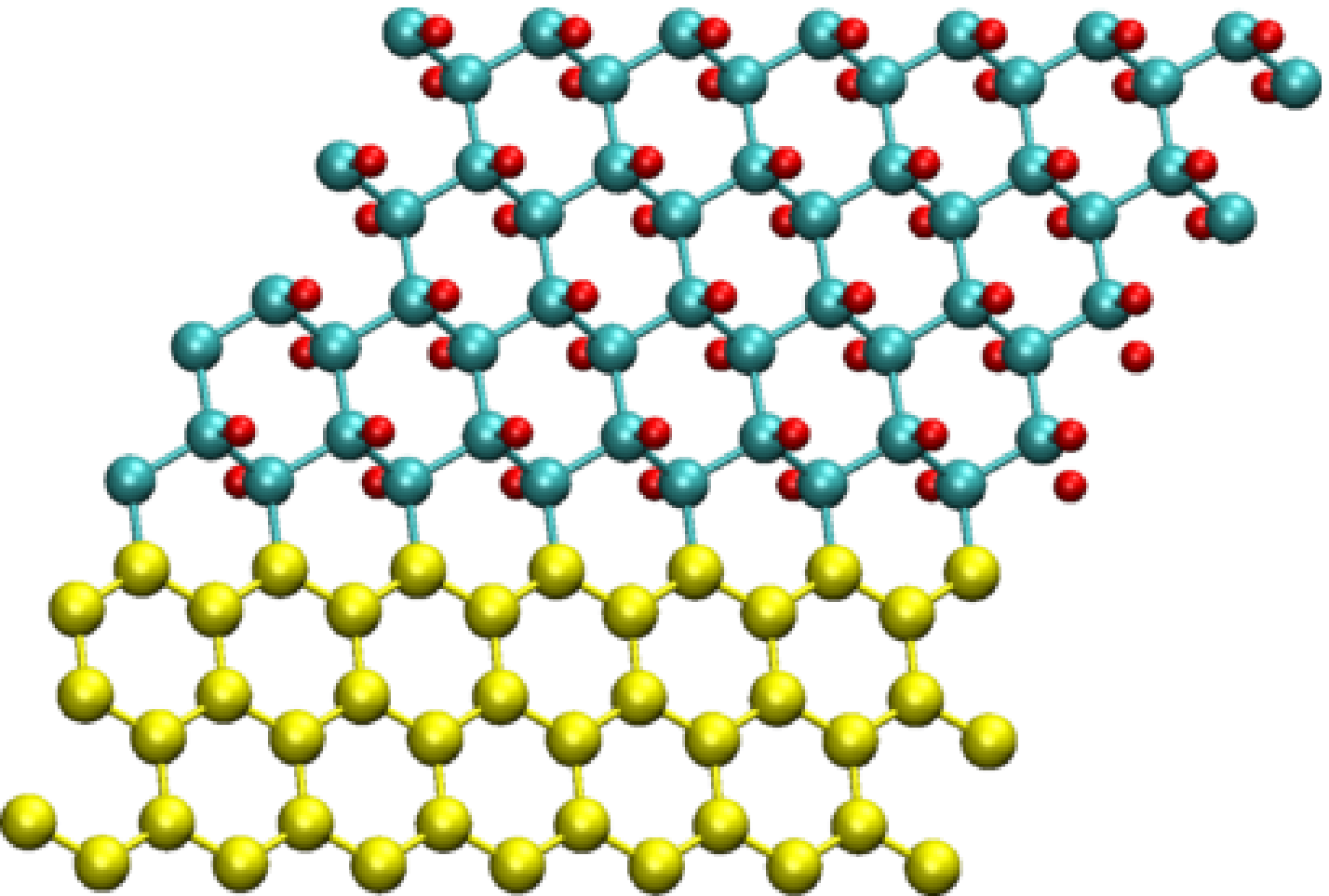} \\
(a) \hskip 5cm (b) \\~\\
\includegraphics[width=8.0cm]{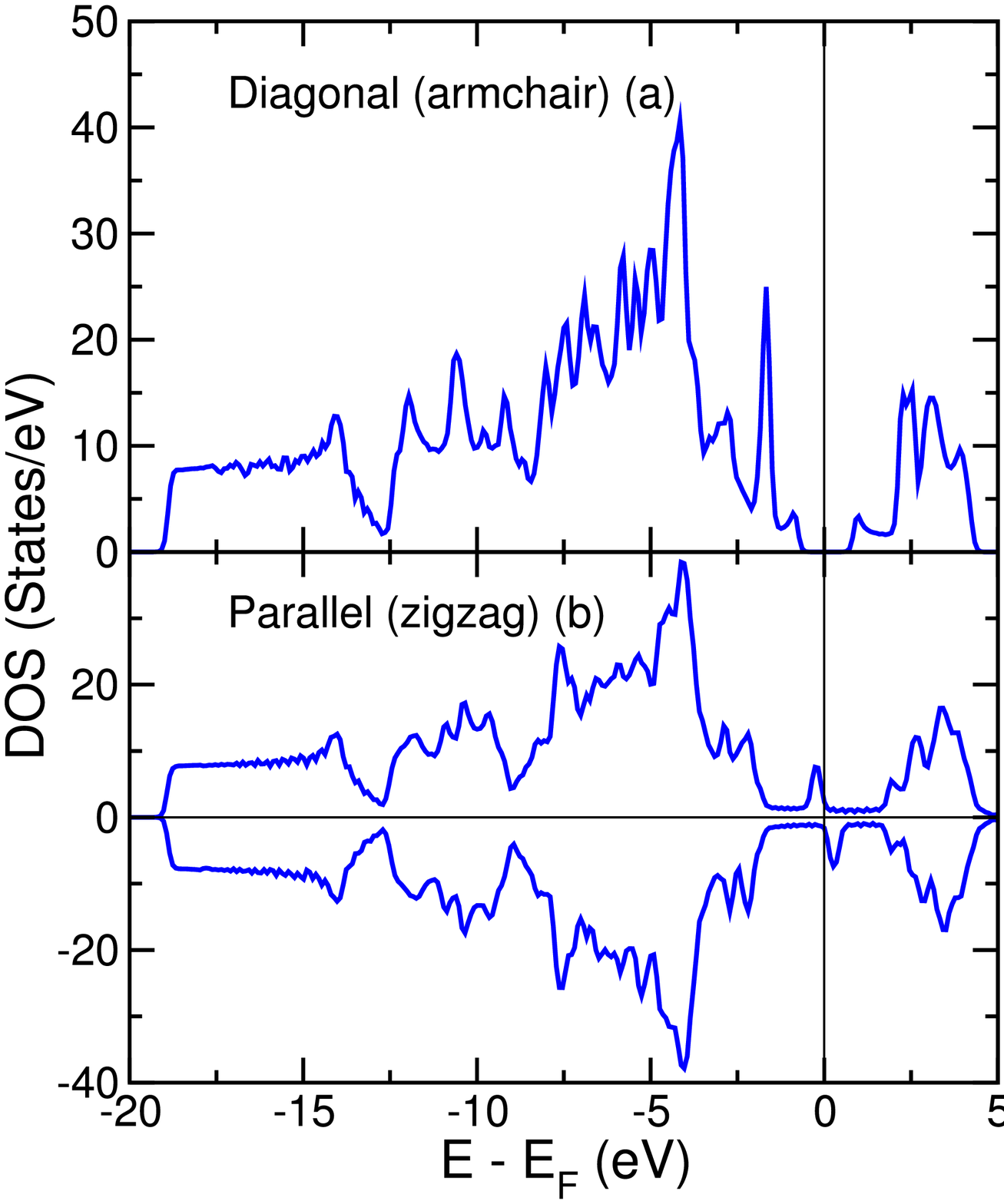} \\
(c)
\end{center}

\caption{\label{fig9} (Colour online) Decoration of hydrogen (a) along the diagonal of the
unit cell and (b) along the edge of the unit cell. In figure, yellow (in print light
shaded) balls are bare carbon atoms, turquoise (in print darker shades) balls are
hydrogenated carbons and red (in print dark  small) balls are hydrogen atoms. (c) DOS
plots for the two cases. The lone atoms seen are from repeating super cells}

\end{figure*}

Now we bring in another interesting aspect of hydrogenation brought out by our
calculations. Our results indicate that it is possible to pattern the graphene lattice
with hydrogen and tune the electronic structure. In figure \ref{fig9}(a), the patterning
is done by removing the hydrogen atoms along the diagonal of the unit cell where as in
figure \ref{fig9}(b) the hydrogen atoms are removed parallel to one edge of the unit cell.
The horizontal pattern is more stable than the diagonal pattern by 0.04 eV/H atom. The
corresponding DOS are shown in figure \ref{fig9}(c) where a dramatic difference is seen.
The diagonal pattern shows a clear band gap of 1.4 eV whereas the horizontal pattern shows
finite DOS along with a magnetic solution. A close inspection of the geometry reveals an
analogy with graphene nano ribbons (GNR). The diagonal pattern resembles an armchair GNR
with a chain of hexagons while the horizontal pattern is analogous to a zigzag GNR with a
width of 3 rows in this particular case. The corresponding DOS also resemble the
electronic structure of an armchair (zigzag) GNR with a semiconducting (metallic)
nature~\cite{Loui}. It is a reasonable conjecture that the band gaps of the patterned
system can be tuned by controlling the width of the bare carbon channels analogous to the
case of armchair GNR where the band gap decreases as the width of GNR is increased.

\section{Concluding Remarks}

In conclusion, our detailed density functional investigations have revealed some novel
features of graphene-{\it graphane} metal-insulator transition.  As the hydrogen coverage
increases, graphene with a semi metallic character turns first into a metal and then to an
insulator. Hydrogenation of graphene pulls the carbon atom out of the plane breaking the
symmetry of pure graphene. As a consequence, many {\bf $k$} points in the Brillouin zone
contribute to the DOS at the Fermi level giving rise to a metallic system. The metallic
phase has some unusual characteristics: the sheet shows two distinct regions, a conducting
region formed by bare carbon atoms and embedded into this region are the non conducting
islands formed by the hydrogenated carbon atoms.  The onslaught of insulating state occurs
when there are insufficient numbers of bare carbon atoms to form connecting channels.
This also means that the transition to insulating phase depends on the distribution of
hydrogen atoms and will occur when the continuous channels are absent.  The present work
opens up the possibility of using partially hydrogenated graphene having designed patterns
of conducting channels along with insulating barriers for the purpose of devices. Our
results also show that it is possible to design a pattern of hydrogenation so as to yield
a semiconducting sheet with a band gap much lower than that of {\it graphane}.  Finally we
may note that the calculation of conductivity in such a disordered system  is a complex
issue. The present study focusses on the evolution of the density of states to understand
the change in the character of single particle orbitals as a function of hydrogen
coverage. An obvious extension of this work is the study of transport properties to have a
more vivid picture.

\section*{Acknowledgement}

S.H. would like to acknowledge Indo-Swiss grant for financial support (No:
INT/SWISS/P-17/2009). B.S.P. would like to acknowledge CSIR, Govt. of India for financial
support (No: 9/137(0458)/2008-EMR-I). B.S. and B.S.P. are grateful to Swedish Research
Council (VR) and Swedish Research Links programme funded by VR/SIDA for financial
assistance and Swedish National Infrastructure for Computing (SNIC) for allocating
supercomputing facilities. B.S. acknowledges financial support from Carl Tryggers
Foundation and KOF initiative of Uppsala University. B.S.P., S.H. and D.G.K. acknowledge
the Swedish Research Links programme for their visits to Uppsala university. Some of the
figures are generated by using VMD~\cite{vmd} and XCrySDen~\cite{xcrysden}. P.C.
acknowledges M. Arjunwadkar for encouragement and discussion.

\bibliographystyle{unsrt}
\bibliography{biblio}

\end{document}